\documentclass[graybox, nosecnum]{svmult}

\usepackage{mathptmx}       
\usepackage{helvet}         
\usepackage{courier}        
\usepackage{type1cm}        
\usepackage{makeidx}         
\usepackage{graphicx}        
\usepackage{multicol}        
\usepackage[bottom]{footmisc}
\usepackage{hyperref}        
\usepackage{soul}            
\hypersetup{colorlinks=true,urlcolor=blue}
\usepackage{amssymb}
\usepackage{newtxmath}
\usepackage[square,numbers]{natbib}
\usepackage{amsmath}
\newcommand{\hbindex}[1]{\hl{#1}\index{#1}}  
\makeindex             
                       
\usepackage{booktabs}

\def\s{{\rm\,s}}
\def\erg{{\rm\,erg}}
\def\km{{\rm\,km}}

\def\msun{${\rm M}_{\odot}$}
\def\chandra{\textit{Chandra~}}

\def\spitzer{\textit{Spitzer}}
\def \ergs {$\rm erg~s^{-1}$}

\def \lx {{L$_{\rm{X}}$}}
\def \lxunit {${\rm erg~s^{-1}}$}
\def \zsun {Z$_{\odot}$}

\def \ha {${\rm H}\alpha$}
\def \pspin {${\rm P_{spin}}$}
\def \porb {${\rm P_{orb}}$}

\begin{document}
\title*{High-mass X-ray Binaries}

\author{Francesca M. Fornasini \thanks{corresponding author}, Vallia Antoniou, and Guillaume Dubus}
\authorrunning{F. M. Fornasini, V. Antoniou, \& G. Dubus}
\institute{F. M. Fornasini \at Stonehill College, 320 Washington Street, North Easton, MA 02357, USA \email{ffornasini@stonehill.edu}
\and V. Antoniou \at Texas Tech University, Department of Physics \& Astronomy, Lubbock, TX 79409-1051, USA \hspace{0.5in} Center for Astrophysics $\vert$ Harvard \& Smithsonian, Cambridge, MA 02138, USA \email{vallia.antoniou@ttu.edu}
\and G. Dubus \at Univ. Grenoble Alpes, CNRS, IPAG, 38000 Grenoble, France \email{guillaume.dubus@univ-grenoble-alpes.fr}}
%
%
\maketitle
\abstract{Binary systems in which a neutron star or black hole accretes material from a high-mass star are known as high-mass X-ray binaries (HMXBs).  This chapter provides a brief introduction to the physics of wind accretion and an observational view of HMXBs, including their classification, X-ray spectra, X-ray variability, orbital and compact object properties, as well as studies of Galactic and Magellanic HMXB populations.  Two classes of X-ray sources whose possible connections to HMXBs have been debated, ultraluminous X-ray sources and gamma-ray binaries, are also discussed.  Approximately 300 HMXBs residing either in the Milky Way or the Magellanic Clouds have been discovered.  The majority of these HMXBs host wind-accreting neutron stars.  Their X-ray properties depend both on the interaction of the accreting material with the neutron star’s strong magnetic field and the properties of the donor star’s wind.  Most HMXBs are classified as either supergiant XBs or Be XBs based on the spectral type of the donor star; these classes exhibit different patterns of X-ray variability and occupy different phase space in diagrams of neutron star spin versus orbital period.  
While studies of HMXBs in the Milky Way and Magellanic Clouds find that their luminosity functions have similar shapes, an overabundance of Be XBs in the Small Magellanic Cloud points to important variations of the HMXB population with metallicity and age.}

\section{Keywords} 
X-ray binaries; high-mass X-ray binaries; high-mass stars; neutron stars; wind accretion; Galactic X-ray sources; Magellanic X-ray sources
\section{1 Introduction}
\label{sec:intro}

\hbindex{High-mass X-ray binaries} (HMXBs) consist of a neutron star (NS) or black hole (BH) accreting matter from a stellar binary companion with a mass $\gtrsim10 M_{\odot}$.  The X-ray emission in these systems is produced as plasma falls toward the compact object, releasing gravitational potential energy and heating up.  Galactic HMXBs were among the first X-ray sources detected by rocket and balloon-borne X-ray detectors in the 1960s (e.g. \cite{giacconi62}) and the first X-ray astronomy satellites in the 1970s (e.g. \cite{giacconi74}).  Over the past six decades, approximately 150 HMXBs have been identified in our Milky Way galaxy \citep{fortin23}, and roughly 200 HMXBs have been discovered in the Large and Small Magellanic Clouds (LMC; \cite{antoniou16}, SMC; \cite{haberl16}).   Populations of HMXBs have been resolved in over 50 nearby galaxies (e.g. \cite{mineo12,lehmer21}) and their integrated emission has been detected out to high redshift (e.g. \cite{lehmer16}). \par
In the majority of HMXBs, the compact object is a \hbindex{neutron star} accreting matter directly from the stellar wind of its binary companion \citep{liu06}.  \hbindex{High-mass stars} have powerful stellar winds with mass loss rates of $\gtrsim10^{-6} M_{\odot}$ yr$^{-1}$ \citep{smith14}, so direct \hbindex{wind accretion} can result in relatively high accretion rates for binary separations of a few AU or less, producing X-ray luminosities in the $10^{35}-10^{40}$ erg s$^{-1}$ range.  The donor star in most HMXBs is either a supergiant O/B star or a Be star, both of which have higher mass loss rates than O/B main sequence stars. \par
HMXBs form through the evolution of binary systems consisting of two high-mass stars.  Most of the binaries that evolve into HMXBs are expected to pass through a phase of mass transfer prior to the supernova explosion of the more massive star, which typically helps to shrink the binary orbit \cite{postnov14}.  The HMXB phase typically begins $\approx4-40$ Myr after the formation of the high-mass binary and lasts $\sim10^4$ years (\cite{linden10,postnov14,antoniou16}).  Given their rapid formation times, the first generations of HMXBs may provide a substantial amount of X-ray heating to intergalactic gas during the Epoch of Reionization, impacting the onset and duration of this major phase transition in the early Universe \cite{jeon14}, which is one of the major next frontiers in astrophysics.  Some HMXB systems may eventually evolve into double compact objects, which may merge and produce gravitational waves \cite{belczynski02}.  Thus, understanding the formation and evolution of HMXBs provides important insights into one of the primary formation channels of gravitational wave sources, a key puzzle in the new field of gravitational wave astrophysics.    

This chapter begins with an overview of accretion physics in HMXBs, focusing on wind accretion, which applies to the majority of HMXBs, but also briefly discussing accretion via Roche-lobe overflow.  Then follows discussion of different classes of HMXBs, which primarily depend on the spectral type of the donor star since the wind properties of the donor significantly impact the observed X-ray properties.  Other classes of sources including gamma-ray binaries and ultraluminous X-ray sources are also discussed.  Next, mass measurements of the compact objects in HMXBs are summarized.  Following this are descriptions of the X-ray spectral properties and the temporal behavior of HMXBs, including X-ray variability, NS spin periods, and binary orbital periods.  Finally, studies of the HMXB populations in the Milky Way, LMC, and SMC are discussed.  For additional information, see recent reviews on HMXBs \cite{walter15,martinez17,kretschmar19} and references therein.

\section{2 Accretion in HMXBs}
\label{sec:accretion}
Among the different types of X-ray bright accreting binaries, HMXBs are fairly unique in that the mass transfer mechanism in most of them is wind accretion rather than Roche lobe overflow (RLO).  A star's Roche lobe is defined as the gravitational equipotential boundary within which material is gravitationally bound to the star.  Roche-lobe overflow occurs when the donor star in a binary expands beyond its Roche lobe and the matter outside its Roche lobe is then gravitationally attracted to the its binary companion; this material flows from the donor star towards the accretor in an accretion stream, forming an accretion disk around the accretor.  Accretion in both low-mass X-ray binaries (LMXBs) and cataclysmic variables occurs via RLO; the winds of the low-mass donor stars in such binaries are too weak to produce significant wind accretion.  Besides HMXBs, another group of binaries in which wind accretion is common is symbiotic binaries, most of which consist of a white dwarf and a red giant companion.  However, symbiotic binaries are much fainter than HMXBs, with typical X-ray luminosities of $L_X\sim10^{30}-10^{33}$ erg s$^{-1}$.  \par
While a few known HMXBs are thought to be approaching or experiencing RLO (i.e. SMC X-1, Cen X-3, LMC X-4), the high mass ratio in these binaries is predicted to make RLO unstable, resulting in a rapid shrinking of the orbit and leading to a common envelope phase \citep{vandenheuvel17}.  In this scenario, the X-ray active phase cannot last longer than the thermal timescale of the donor star, which is $\sim10^4$ yr for an O star \citep{kretschmar19}, which explains the scarcity of known RLO HMXBs.  
While RLO HMXBs are rare, it is not unusual for temporary accretion disks to form around the compact objects in HMXBs while they are accreting from their companion's stellar wind.  We discuss the situations in which accretion disks form in HMXBs in \S\hyperref[sec:diskfed]{2.1} and provide an overview of wind accretion in HMXBs in \S\hyperref[sec:windaccretion]{2.2}.

\subsection{2.1 Disk-fed accretion}
\label{sec:diskfed}
The presence of an \hbindex{accretion disk} in some HMXBs has been inferred from different observational features including:  (i) a high luminosity and corresponding high accretion rate above the maximum expected by direct wind accretion, (ii) their position in the lower left quadrant of the spin period versus orbital period (Corbet) diagram \citep{corbet86} corresponding to low spin and orbital periods ($P_{\mathrm{spin}}\lesssim10$ seconds, $P_{\mathrm{orb}}\lesssim5$ days), (iii) rapid spin-up rates indicating that the material being accreted has coherent angular momentum (e.g. \cite{jenke12}), (iv) detailed modeling of their optical light curves requiring the presence of an accretion disk component \citep{tjemkes86}, or (v) optical spectra revealing the presence of lines such as HeII 4686 \AA\ that require hotter temperatures than the surface temperatures of supergiants and have radial velocity curves in anti-phase with the donor star lines \citep{hutchings77}. Examples of HMXBs which exhibit strong evidence for accretion disks are SMC X-1, Cen X-3, and LMC X-4, which are thought to be experiencing RLO, as well as Cyg X-1, whose orbital properties have been measured well enough to know that the system is close to but not currently experiencing RLO \citep{kretschmar19}.  \par
Transient disks have been reported in other HMXBs with supergiant donor stars, including OAO 1657-415, 4U 0114+650, GX 301-2, IGR J08408-4503, and Vela X-1 \citep{kretschmar19,liao20}.  Studies have found that in HMXBs with close binary separations or in which the supergiant donors have slower wind speeds than are typical for isolated supergiant OB stars (see \S\hyperref[sec:sghmxb]{3.2}), the stellar wind can be focused in the direction of the compact object \citep{friend82} and accretion disks can temporarily form in wind-accreting systems not undergoing RLO \citep{kretschmar19}.  These models have been successfully applied to explain the formation of accretion disks in HMXBs including Cyg X-1, Vela X-1, and OAO 1657-415, and may help to explain the transient disks seen in some of other supergiant HMXBs.  
\par
As discussed in \S\hyperref[sec:bexrb]{3.3}, accretion disks also tend to form during outbursts in HMXBs with Be donor stars.  The rapid spin-up rates of neutron stars in Be X-ray binaries (Be XBs) observed consistently during Type II outbursts and during some Type I outbursts indicate that accretion disks are present during Type II and some Type I outbursts \citep{ziolkowski02}.  It has also been suggested that, at least in some Be XBs, a cold (low-ionization) accretion disk may persist while the source is in a quiescent state \citep{mushtukov22}. \par
Accretion disks are also present in ultraluminous X-ray sources (ULXs), many of which are X-ray binaries in which the compact object is accreting at super-Eddington rates, as detailed in \S\hyperref[sec:ulx]{3.5}.  The spectra of many ULXs show evidence for powerful outflows and geometrically thick accretion disks, signatures of super-Eddington accretion.  The high accretion rates observed in ULXs may result from RLO, although some simulations have shown that highly focused winds in close binaries not undergoing RLO may produce sufficient mass transfer rates to power ULXs \citep{elmellah19b}.  

\subsection{2.2 Wind accretion} 
\label{sec:windaccretion}
\begin{figure}[t]
\centering
\includegraphics[width=0.65\textwidth, angle=90]{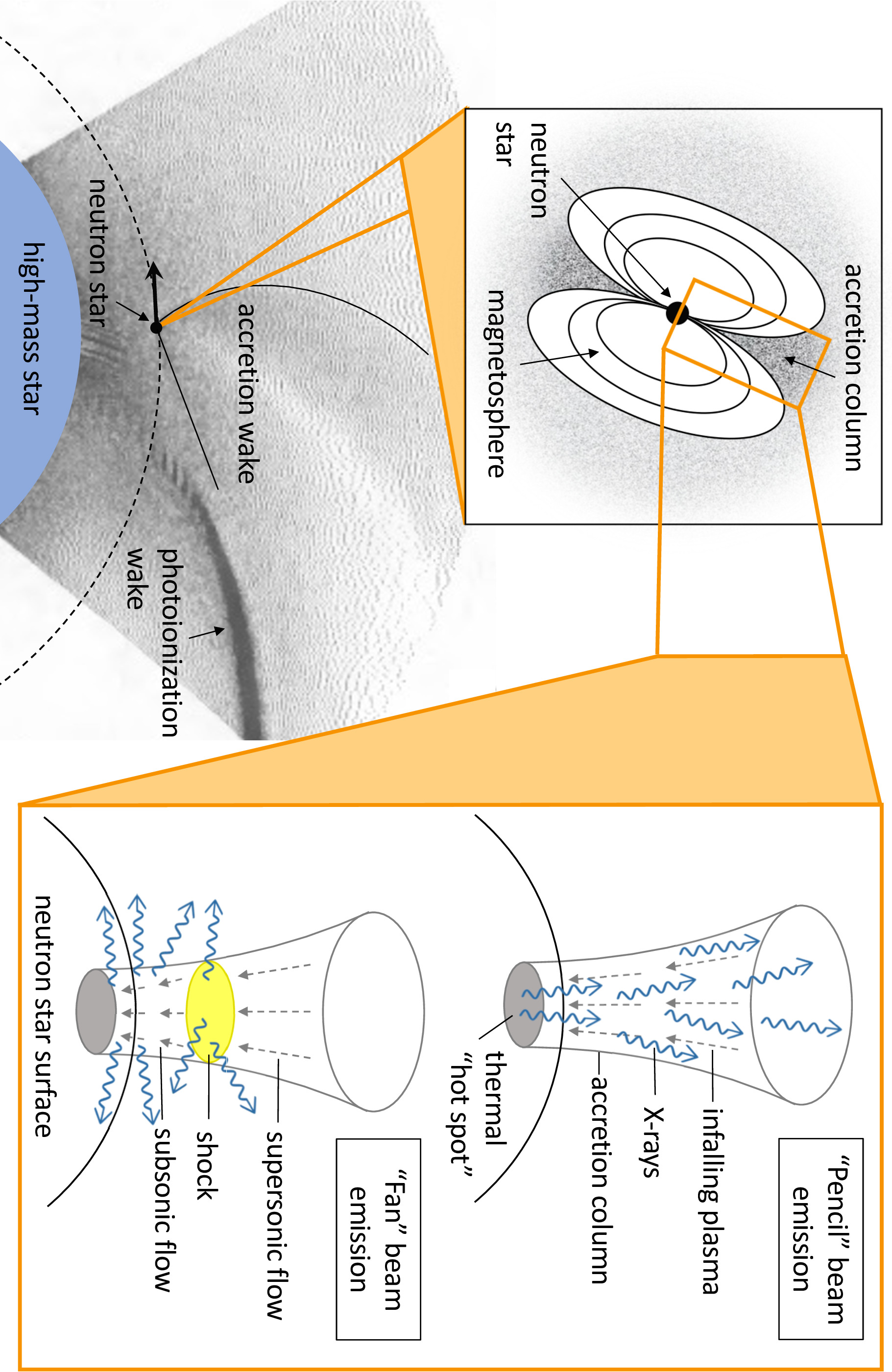}
\caption{Schematic view of the accretion geometry in a typical HMXB.  \textit{Lower left:} Simulation of the stellar wind density by \cite{blondin90} showing the formation of an accretion wake and a photoionization wake due to the orbital motion and X-ray emission of the accreting neutron star.  See \S\hyperref[sec:classicalhmxb]{3.2.1} for discussion of the formation of these wakes.  \textit{Upper left:} The strong magnetic field of the neutron star channels the inflowing matter (shown in gray) into accretion columns above the magnetic polar caps. \textit{Right:} Diagrams of possible X-ray emission geometries from the accretion column.  Pencil beam emission is produced when X-rays produced by the thermal ``hot spot" at the NS surface or within the accretion column can mostly stream freely through the infalling plasma.  Fan beam emission results when X-rays are efficiently scattered out of the accretion column by the infalling plasma, which can occur, for example, when a shock forms within the column.}
\label{fig:hmxbschematic}       
\end{figure}

The vast majority of HMXBs are wind accretors.  Most contain neutron stars, whose strong magnetic fields channel the accretion flow onto the magnetic poles in the vicinity of the NS, as shown in Fig. \ref{fig:hmxbschematic}.  Since a significant fraction of the X-ray emission is produced in regions near the magnetic poles (see \S\hyperref[sec:spectra]{5.1}), when the magnetic and rotation axes are misaligned, X-ray pulsations can be observed as the NS rotates (see \S\hyperref[sec:pulsations]{6.3.1}).  Such pulsating X-ray sources are referred to as \hbindex{pulsars}.  Although the accretion flow in the vicinity of the NS is strongly affected by its magnetic field, and in some cases an accretion disk may form around the NS (see \S\ref{sec:diskfed}), the accretion rate onto the NS can be approximately estimated using simple spherical accretion models.  \par
Thus, the X-ray luminosities of persistent, wind-fed HMXBs can be explained using a simple model for spherically symmetric accretion.  The X-ray luminosity produced is, to order-of-magnitude, equal to the gravitational potential energy released by the infalling material per unit time, which can be expressed as:
\begin{equation}
L_X\approx\frac{GM_c\dot{M}_c}{R_c}\approx0.1\dot{M}_c c^2
\end{equation}
where $M_c$  and $R_c$ represent the mass and radius of the compact object, respectively, and $\dot{M}_c$ is the mass accretion rate.  The mass accretion rate onto the compact object is $\dot{M}_c \approx \pi R_B^2 \rho \varv$, where $\rho$ and $\varv$ are the wind density and velocity, respectively, at the position of the compact object, and $R_B$ is the radius within which the wind material is gravitationally captured by the compact object \citep{bondi44}.  This radius, often referred to as the Bondi-Hoyle-Littleton radius, is defined as the location where the wind velocity equals the escape velocity of the compact object:
\begin{eqnarray}
\varv &=& \sqrt{\frac{2GM_c}{R_B}} \nonumber \\
R_B &=& \frac{2GM_c}{\varv^2} 
\end{eqnarray}
Thus, the accretion rate onto the compact object is:
\begin{equation}
\dot{M}_c \approx 4\pi G^2M_c^2\rho/\varv^3.
\end{equation}  \par
The wind velocity as a function of distance from the donor star, $r$ can be parametrized with the $\beta$-velocity law \citep{castor75}:
\begin{equation}
\varv(r)=\varv_{\infty} (1-R_*/r)^\beta
\end{equation}
where $\varv_{\infty}$ is the terminal velocity of the wind at large distances from the star, and $R_*$ is the stellar radius.  For O and B type supergiants, which make up the majority of donor stars in persistent wind-fed HMXBs, the terminal wind speeds are $\varv_{\infty}\sim500-2000$ km s$^{-1}$ and typical values of $\beta$ are between 0.5 and 1 \citep{kudritzki00}.  The stellar radius of OB supergiants depends on their stellar mass, roughly following the relationship from \cite{lutovinov13}:
\begin{equation}
\frac{R_*}{R_{\odot}} \approx 0.9\frac{M_*}{M_{\odot}}.
\end{equation}
\par
The density of the stellar wind is related to the mass loss rate from the donor star.  For a spherically symmetric stellar wind, the mass loss rate is $\dot{M}_w = 4\pi r^2 \rho(r) \varv(r)$.  For the radiatively-driven winds of high-mass stars, the momentum of the stellar wind is comparable to the momentum of the stellar radiation.  Therefore, $\dot{M}_w \varv_{\infty} \approx \epsilon L/c$, where $\epsilon$ represents the efficiency of momentum transfer from the star's radiation to the wind material and is typically equal to $0.4-1.0$ for OB supergiants \citep{lutovinov13}.  Combining these equations, the density of the stellar wind can be expressed as:
\begin{equation}
\rho(r) \approx \frac{\epsilon L}{4\pi r^2 \varv(r) \varv_{\infty} c}.
\end{equation}
The luminosity of OB supergiants primarily depends on their stellar mass, following the relationship from \cite{vitrichenko07}:
\begin{equation}
\frac{L}{L_{\odot}} \approx 19\left(\frac{M_*}{M_{\odot}}\right)^{2.76}.
\end{equation}
\par
Combining Eqs. 1-7 and evaluating the wind parameters for the semi-major axis of the binary orbit ($r=a$), results in a relationship between $L_X$, $M_*$, $M_c$, $\varv_{\infty}$, and $a$:
\begin{multline}
L_X \approx 5\times10^{35} \erg \s^{-1} \left(\frac{M_*}{10 M_{\odot}}\right)^{2.76}\left(\frac{M_c}{1.4 M_{\odot}}\right)^2 \left(\frac{a}{10 R_{\odot}}\right)^{-2} \\ \times \left(\frac{\varv_{\infty}}{1000 \km \s^{-1}}\right)^{-5}\left(1-0.9\frac{M_*}{M_{\odot}}\frac{R_{\odot}}{a}\right)^{-4\beta}
\end{multline}
Given that $M_*$, $M_c$, and $\varv_{\infty}$ occupy a relatively narrow range of possible values for OB supergiant donor stars, this implies that for a given orbital separation, there is a minimum X-ray luminosity for persistent, wind-fed HMXBs, which can be estimated using Eq. 8.  Since by Kepler's law, the semi-major axis is related to the orbital period by $a\propto P_{\mathrm{orb}}^{2/3}$, the minimum X-ray luminosity for wind-fed HMXBs scales as $L_X\propto P_{\mathrm{orb}}^{-4/3}$.  There is evidence that persistent HMXBs do indeed occupy a region of $L_X-P_{\mathrm{orb}}$ parameter space above this relation \citep{lutovinov13}. \par
Modeling the X-ray luminosities of highly variable and transient HMXBs, many of which host Be donor stars, requires the inclusion of more complex factors, such as the inhomogeneous structure of stellar winds, eccentric orbits, and the impact of the NS magnetosphere on the accretion flow.  

\subsection{2.3 Interactions between the accretion flow and the magnetosphere} 
\label{sec:magnetosphere}
\begin{figure}[t]
\centering
\includegraphics[width=1.0\textwidth]{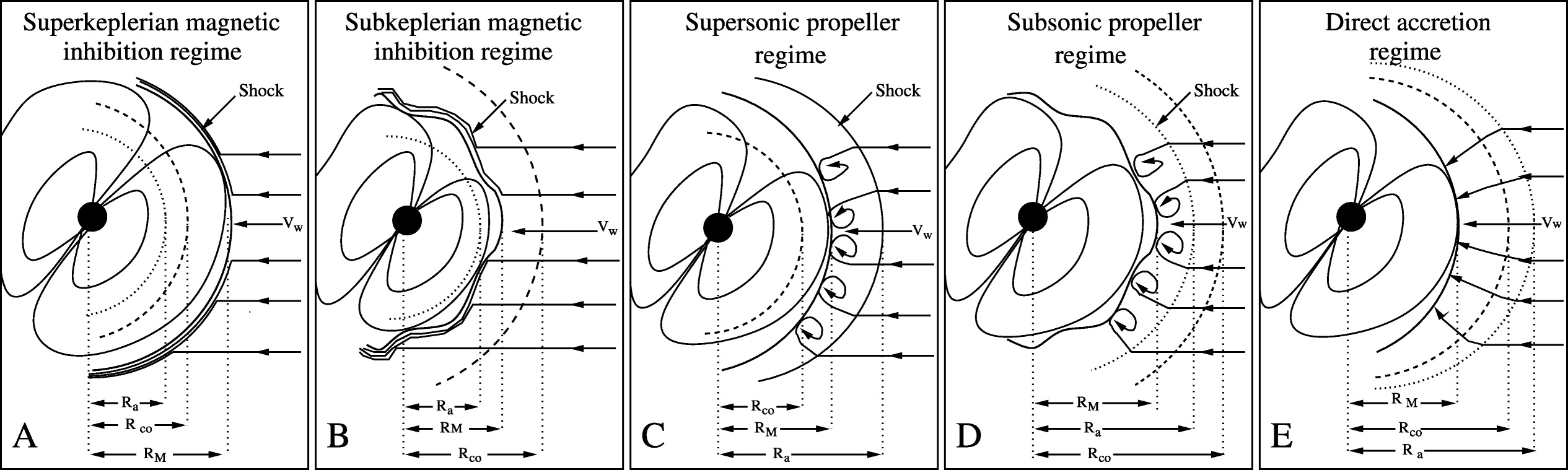}
\caption{Schematic view of a magnetized NS interacting with the inflowing matter from its high-mass stellar companion. Different accretion regimes are shown, together with the relative position of the magnetospheric radius $R_M$ (solid line), the corotation radius $R_{\mathrm{co}}$ (dashed line), and the accretion radius $R_{\mathrm{a}}$(dotted line). A wavy solid line represents when the magnetospheric boundary at $R_{M}$ is Kelvin-Helmholtz unstable. In the supersonic and subsonic propeller regimes, convective motions at the base of the quasi-spherical atmosphere are represented by small eddies. Credit: Figure 1 in \cite{bozzo08}, reproduced by permission of the AAS.}
\label{fig:bozzo08}       
\end{figure}
All HMXBs exhibit variability and some exhibit such strong variability that they are transient sources.  Some of this variability and transient behavior is associated with the inhomogeneity and detailed structure of the stellar winds, as described in \S\hyperref[sec:classes]{3}.  \par
Since the majority of HMXBs contain neutron stars, additional variability can be introduced by the interaction of the accretion flow with the \hbindex{magnetosphere}, which rotates with the same angular velocity as the neutron star.  The magnetospheric radius is defined by the balance between magnetic pressure [$B^2/(8\pi)\mu^2/(8\pi R^6)$, where $\mu$ is the magnetic moment and $R$ is the distance from the center of the neutron star, assuming a dipolar field] and the ram pressure of the accreting material ($\rho \varv^2$).  In the case that the accreting matter free-falls towards the neutron star surface, the matter density can be related to the mass accretion rate onto the neutron star through $\rho = \dot{M}_c/(4\pi R^2 \varv)$, and its velocity is given by $\varv = \sqrt{2GM_c/R}$.  Therefore, in the free-fall case, the magnetospheric radius is:
\begin{equation}
R_M = \left(\frac{\mu^2}{\dot{M}_c\sqrt{2 GM_c}}\right)^{2/7}
\end{equation}
The equation for the magnetospheric radius may differ in certain situations, for example if the accreting material is not in free fall, but instead its bulk motion is determined by the stellar wind velocity at the location of the neutron star.  \par
As shown in Fig. \ref{fig:bozzo08} and described in \cite{bozzo08}, different accretion regimes produced by interactions between the accretion flow and the magnetosphere can be defined by considering three radii: the magnetospheric radius ($R_M$), the corotation radius ($R_{\mathrm{co}}$) defined as the location where the angular velocity of the neutron star equals the Keplerian angular velocity, and the accretion (or Bondi) radius ($R_{\mathrm{a}}$) defined at the distance where the inflowing matter is gravitationally focused towards the neutron star.  Depending on the relationship between these radii, accretion can be magnetically or centrifugally inhibited.  If $R_M>R_{\mathrm{a}}$, the accreting material forms a bow shock as it runs into the magnetosphere, resulting in a magnetic barrier or ``gate" to accretion that can halt or significantly reduce the accretion rate onto the NS.  If $R_{\mathrm{M}}>R_{\mathrm{co}}$, the rotational velocity of the magnetosphere is supersonic relative to the infalling material, resulting in a centrifugal barrier to accretion often referred to as the propeller effect.  In this case, since the accreting material has a lower rotational velocity than the magnetosphere, it exerts a torque on the magnetosphere, dissipating some of the neutron star's rotational energy, thus causing a spin-down of the star.  Direct accretion occurring when $R_{\mathrm{a}}>R_{\mathrm{co}}>R_M$ can spin-up the NS, especially if an accretion disk is formed.
Magnetospheric barriers to accretion seem required to explain the behavior of supergiant fast X-ray transients, a class of HMXBs described in Section \hyperref[sec:SFXT]{3.2.3}, and may contribute to the variability of other HMXBs as well.  For more details on the effects of the magnetosphere on wind accretion see \cite{bozzo08, kretschmar19, mushtukov22}.  

\section{3 Classes of High-Mass X-ray Binaries}
\label{sec:classes}
Since the compact objects in the majority of HMXBs accrete matter from the stellar wind of their binary companions, their X-ray properties strongly depend on the properties of the donor stellar wind.  Therefore, HMXBs are often classified based on the spectral type of the donor star, as well as the type of compact object they host.  \par
\begin{figure}[t]
\centering
\includegraphics[width=0.4\textwidth,angle=90]{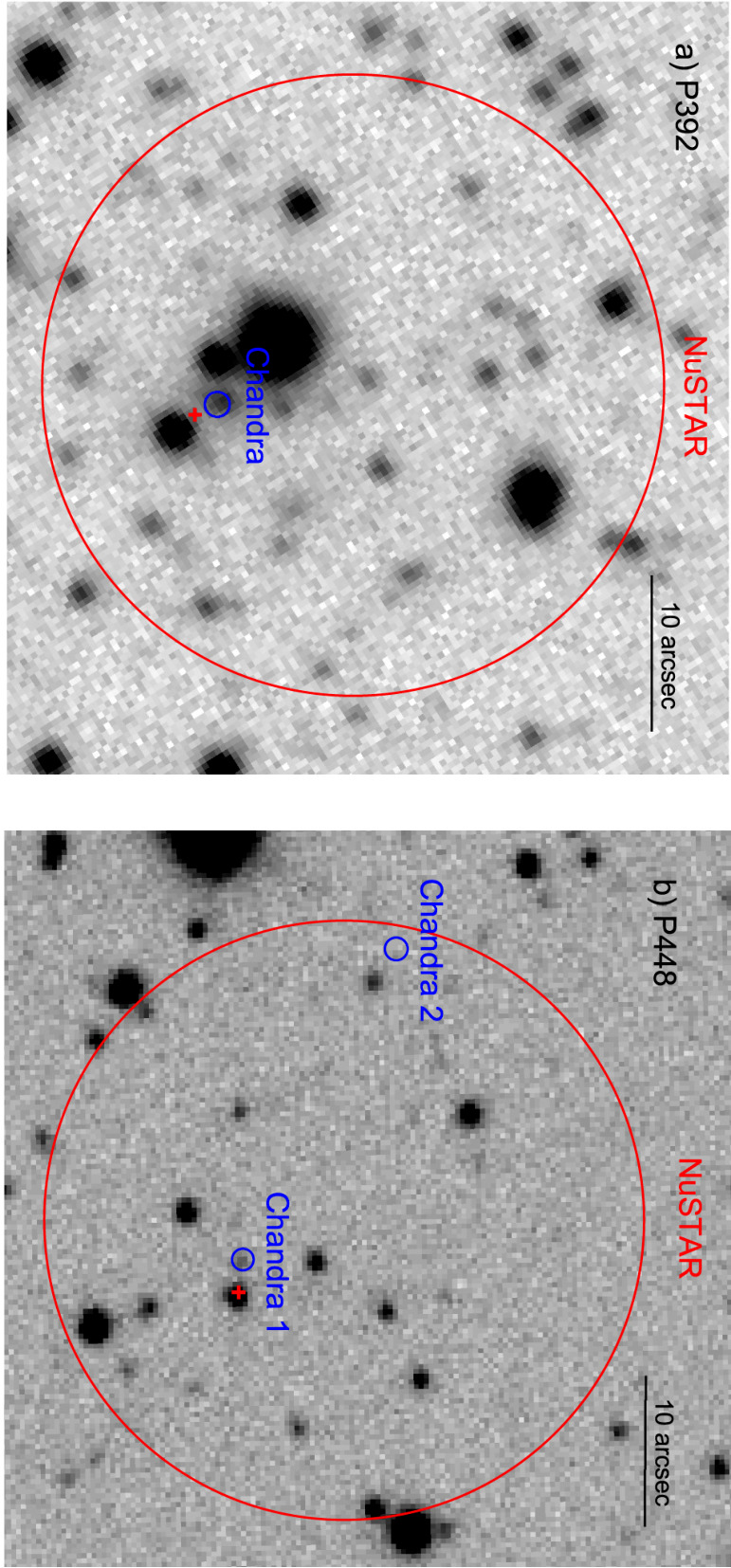}
\caption{Example of the multi-wavelength process required to classify new hard X-ray sources.  In each optical image, the error circle of a source discovered by \textit{NuSTAR} near the Galactic plane is shown in red, and the error circles of soft X-ray sources detected in follow-up \textit{Chandra} observations are shown in blue, numbered from brightest to faintest.  The i-band images come from (a) the Sloan Digitized Sky Survey, and (b) the IPHAS survey. The red + symbols mark the positions of the most likely optical counterparts determined by a previous study based on available soft X-ray and optical catalogs; the soft X-ray catalogs tend to have larger positional uncertainties compared to the on-axis follow-up \textit{Chandra} observations.  As can be seen, for hard X-ray sources in crowded Galactic fields, \textit{Chandra}'s positional accuracy is crucial for identifying the correct optical counterpart.  Credit: Taken from Figure 4 in \cite{tomsick18}, reproduced with permission \textcopyright AAS.}
\label{fig:tomsick18}       
\end{figure}

While hard X-ray telescopes (i.e. \textit{INTEGRAL}, \textit{NuSTAR}) have been very effective at discovering HMXBs, classifying HMXBs requires identifying their optical/infrared counterparts.  Since the angular resolution of hard X-ray telescopes ($\sim0.5-5$ arcminutes) is often insufficient for identifying unique optical/infrared counterparts, it is often necessary to first more precisely determine the source position using a soft-X-ray telescope (i.e. \textit{XMM-Newton}, \textit{Chandra}), as exemplified in Fig. \ref{fig:tomsick18}.  For HMXBs in crowded Galactic fields or in external galaxies that have many  possible optical counterparts even after being localized with \textit{Chandra}, the chance coincidence probability of all possible counterparts identified within a search radius based on the X-ray positional uncertainty can be calculated to identify the most likely counterpart.  The spectral type of the optical/infrared counterpart can then be determined via optical/infrared spectroscopy, or in cases when spectroscopy is not feasible, color–magnitude and/or color–color diagrams can be used for an approximate classification.  \par
The donor stars in most HMXBs have been determined to be either supergiant (Sg) O/B stars or Be stars.  The properties of these two primary classes of HMXBs, Sg XBs and Be XBs, are described first.  Wolf-Rayet X-ray binaries and two additional classes of X-ray sources with possible connections to HMXBs, ultraluminous X-ray sources and gamma-ray binaries, are also presented.  Finally, methods for distinguishing between NS and BH HMXBs are discussed.

\subsection{3.1 Supergiant X-ray Binaries}
\label{sec:sghmxb}
Supergiant (Sg) XBs are primarily wind-fed systems with supergiant O or B type (luminosity class I-II) stars.  The winds of these high-mass stars are driven by the absorption of ultraviolet resonance lines, and the extent of the acceleration depends on the ionization, excitation, and chemical composition of the stellar wind \citep{kudritzki00}.  The ionization of the wind material is affected both by the effective temperature of the star and the X-ray emission from the accreting compact object \citep{kretschmar19}.  The wind mass loss rates of OB supergiants are typically in the range of $10^{-6}-10^{-5} M_{\odot}$ yr$^{-1}$, several orders of magnitude higher compared to the $10^{-10}-10^{-7} M_{\odot}$ yr$^{-1}$ mass loss rates of main sequence O and B stars \citep{smith14}.  \par
Our census of Sg XBs in the Milky Way Galaxy and our understanding of their properties changed substantially with the launch of \textit{INTEGRAL} in 2002.  Due to its improved sensitivity at hard X-ray energies ($17-100$ keV) and its repeated surveys of the Galactic plane, \textit{INTEGRAL} has discovered 40 new confirmed HMXBs \citep{krivonos12, krivonos22}, more than half of which are Sg XBs located in our Galaxy \citep{walter15}.  Prior to \textit{INTEGRAL}, only about 10\% of the roughly 130 HMXBs known were classified as Sg XBs \citep{liu00}, so the prevalence of Sg XBs among the new \textit{INTEGRAL} HMXBs was a surprise.  Most of the \textit{INTEGRAL} Sg XBs are either highly obscured or extremely variable, explaining their late discovery through monitoring, hard X-ray observations.  
About one third of the HMXBs detected in the Milky Way Galaxy are now classified as Sg XBs \citep{walter15,fortin23}, but the vast majority ($>80$\%) of HMXBs in the LMC and SMC are Be HMXBs \citep{haberl16,antoniou16}, which are described in \S\hyperref[sec:bexrb]{3.2}.  

\subsubsection{3.2.1 Persistent ``Classical" HMXBs}
\label{sec:classicalhmxb}
Classical Sg XBs are persistent sources with luminosities consistently above $\sim10^{35}$ erg s$^{-1}$.  However, their luminosity is not constant with time; their flux typically varies by a factor of $10-100$ \citep{walter15,kretschmar19}.  Most Sg XBs are wind-fed systems, with typical orbital periods of $3-60$ days, and most host NSs with spin periods $\gtrsim100$ seconds.  Some of their variability may be due to the clumpy nature of the winds of massive stars.  While there is observational evidence of clumping in the winds of massive stars, the sizes and masses of the clumps are not well constrained \citep{smith14}.  The fact that line-driven winds are likely to be unstable was recognized by analytical studies in the 1970s, and, more recently, numerical hydrodynamical simulations have found that the line-driven instability can result in density variations up to a factor of $10^4$ \citep{kretschmar19}.  As detailed in \cite{martinez17}, the line-driven instability can account for a lot of the variability observed in classical HMXBs.  However, the clump masses required to explain some of the X-ray flares seen in some classical Sg XBs, including Vela X-1, 4U 1700-37, and IGR J16418-4532, are a factor of $10-1000$ above the maximum clump mass of $\sim10^{18}$ g predicted by simulations of the line-driven instability.  This may be due to the limitations of current simulations or it may indicate that some of the X-ray variability of these sources cannot be solely attributed to variations in the stellar wind density but may result from the centrifugal or magnetic barriers to accretion due to the magnetosphere described in \S\hyperref[sec:magnetosphere]{2.1.1}. \par
An additional factor which can impact the X-ray variability of Sg XBs is the influence of the accreting compact object on the stellar wind structure.  The X-ray emission is expected to photoionize the gas in the vicinity of the compact object, decelerating the wind due to the decreased availability of resonance lines \citep{walter15}.  This effect has been observed in some Sg XBs, including Vela X-1, GX 301-2, 4U 1907+09, and EXO 1722-363, in which the terminal wind velocities are measured to be lower than expected based on the luminosities of the donor stars.  As shown in Fig. \ref{fig:hmxbschematic}, at high X-ray luminosities, a photoionization wake of high-density material is expected to be produced as the higher velocity wind from the stellar surface runs into the slower wind that has been photoionized by the compact object (e.g. \cite{blondin90}).  Furthermore, the orbital motion of the accreting object can lead to the formation of a bow shock and a trailing accretion wake of low-density, highly-ionized gas, which can also be seen in Fig. \ref{fig:hmxbschematic}.  These impacts on the stellar wind depend on the X-ray luminosity of the compact object, and can introduce additional X-ray variability. \par
\begin{figure}[t]
\centering
\includegraphics[width=0.24\textwidth, angle=90]{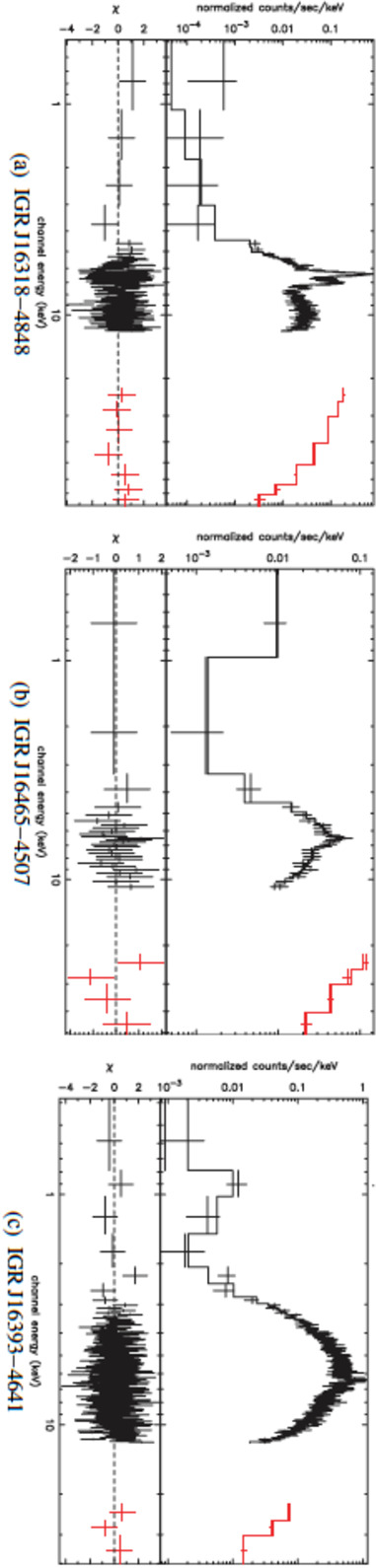}
\caption{X-ray spectra of three highly obscured HMXBs.  Data from \textit{INTEGRAL} ISGRI ($>20$ keV) and \textit{XMM-Newton} EPIC ($<15$ keV) are shown in red and black, respectively.  Lines in the top panels show the best-fitting spectral model for each source and the bottom panels show the data residuals. The spectral model consists of a thermal inverse Comptonization continuum obscured by absorption intrinsic to the source, Gaussian Fe K$\alpha$ and K$\beta$ lines, and a blackbody subject to Galactic absorption along the line-of-sight to represent the soft excess observed in some systems. The sources are ordered from highest ($N_{\mathrm{H}}\approx1.9\times10^{24}$ cm$^{-2}$) to lowest intrinsic absorption ($N_{\mathrm{H}}\approx2.4\times10^{23}$ cm$^{-2}$). Credit: Taken from Figure 4 in Walter, R., et al., A\&A, 453, 133, 2006 \cite{walter06}, reproduced with permission \textcopyright\ ESO.}
\label{fig:walter06}       
\end{figure}
At least six Sg XBs show persistent levels of high obscuration, with hydrogen column densities $N_{\mathrm{H}}\gtrsim10^{23}$ cm$^{-2}$ \citep{walter15}.  The X-ray spectra of three of these HMXBs are shown in Fig. \ref{fig:walter06}; the dearth of X-ray photons below about 4 keV in these spectra is the result of the high obscuring column densities in these systems.  
Most of these highly obscured HMXBs remained undetected until \textit{INTEGRAL}'s sensitive hard X-ray surveys.  Some obscured HMXBs (IGR J16393-4611, IGR J16418-4532, IGR J18027-2016) have short orbital period ($P_{\mathrm{orb}}< 5$ days); such systems may be transitioning from being wind-accreting to Roche lobe overflow systems.   The absorbing column density of one obscured HMXB, EXO 1722-363, varies with orbital phase, which is thought to result from strong perturbations of the stellar wind caused by the massive neutron star in this system.  The high obscuration of IGR J16320-4751, which is fairly constant, may instead be due to the presence of large amounts of dust in the vicinity of the binary given the high infrared reddening observed.  IGR J16318-4848, the HMXB exhibiting the highest obscuration (shown in Fig. \ref{fig:walter06}a), may be orbiting within the dense equatorial outflow of its B[e] supergiant\footnotetext{B[e] supergiants display strong Balmer emission lines and narrow permitted and forbidden emission in the optical band, broad blueshifted resonance lines associated with highly ionized elements in the ultraviolet, and excess emission in the near-infrared band associated with hot circumstellar dust.  The UV lines are thought to originate in a hot, fast polar wind, while the optical lines and near-IR excess are thought to be associated with a cool, slow equatorial wind.} companion \citep{walter06}, because the X-ray absorption is much greater than that of the infrared counterpart, indicating it must be very local to the compact object.  Four additional HMXBs with supergiant B[e] donors have been identified \citep{fortin23}, and at least one of them, CI Cam/XTE J0421+560, is also highly obscured in the X-ray band ($N_{\mathrm{H}}\approx4\times10^{23}$ cm$^{-2}$) \citep{bartlett13}.  While the donor stars in HMXBs are typically identified through optical spectroscopy, the donor stars in highly obscured HMXB systems often require infrared spectroscopy since in many cases the donor stars are also subject to high obscuration and therefore their optical light is highly absorbed.  

\subsubsection{3.2.2 Supergiant Fast X-ray Transients}
\label{sec:SFXT}
Supergiant fast X-ray \hbindex{transients} (SFXTs) were recognized as a new subclass of HMXBs early during INTEGRAL's mission due to their transient properties.  SFXTs exhibit extreme flaring behavior; the flares typically last tens of minutes to a few hours, reaching peak luminosities of $10^{35}-10^{37}$ erg s$^{-1}$ (\cite{sidoli18,kretschmar19}).  Roughly 20 confirmed and candidate SFXTs have been discovered thus far \citep{kretschmar19}.  Several sources displaying behavior that is intermediate between that of SFXTs and classical Sg XBs have also been identified (\cite{walter15,sidoli18}).  \par
SFXTs spend less than 5\% of their time in high-luminosity flaring states; these short flares occur during day-long outbursts during which the SFXT flux is brighter than average \citep{walter15}.  The vast majority of the time, the X-ray fluxes of SFXTs are a factor of $\sim100-10,000$ fainter than the fluxes of their peak flares (\cite{walter15, kretschmar19}).  Typical variability amplitudes are between $L_X\sim10^{33}-10^{34}$ erg/s and $L_X\sim10^{36}$ erg s$^{-1}$ \citep{sidoli18}, although more extreme dynamic ranges from quiescent states at $L_X\sim10^{32}$ erg s$^{-1}$ and flares with $L_X\sim10^{37}$ erg s$^{-1}$ have been observed \citep{walter15}.  SFXTs are so faint in their low flux states that dedicated observations  by sensitive soft X-ray telescopes including \textit{Chandra}, \textit{XMM-Newton}, and \textit{Swift/XRT} are required to study them.  Precise localization by these soft X-ray telescopes allowed the identification of their optical/infrared counterparts, which follow-up spectroscopy revealed to be O or B supergiants (e.g. \cite{masetti06, tomsick08}). \par
While the variability of SFXTs is more extreme than that of other Sg XBs, many of their other properties are similar to to those of other HMXBs.  Based on the fact that X-ray pulsations (see \S\hyperref[sec:pulsations]{6.3.1}) have been detected from a few SFXTs and that other SFXTs have X-ray spectra typical of neutron stars (see \S\hyperref[sec:spectra]{5.1}), it has been determined that most SFXTs harbor neutron stars, just like the vast majority of other HMXBs \citep{kretschmar19}.  While it was proposed that the temporal behavior of SFXTs may indicate they host magnetars\footnote{Magnetars are neutron stars with very strong magnetic fields ($B\gtrsim10^{13}$ G).} \citep{bozzo08}, the few SFXTs for which tentative magnetic field measurements have been made based on spectral cyclotron resonance scattering features (see \S\hyperref[sec:cyclotron]{5.1.1}) are found to host neutron stars with fairly typical surface magnetic field strengths of $B\sim10^{11}-10^{12}$ G \citep{martinez17}.  As discussed in \S\hyperref[sec:corbet]{6.3.3}, the spin periods and orbital periods of SFXTs span similar ranges of values as Sg XBs and Be HMXBs \citep{sidoli18}.  While the orbits of some SFXTs have high eccentricity, others do not \citep{sidoli18}, and there is no clear correlation between orbital phase and SFXT flares \citep{martinez17}.
Therefore, the unusual properties of SFXTs cannot be explained based on their orbital properties. \par
It has been suggested that the extreme variability of SFXTs may be due to these sources being an early stage of Sg XBs, in which the donor stars are more compact and their winds more structured \citep{chaty16}.  There is observational evidence that the stellar winds of SFXTs differ from those of classical Sg XBs.  Overall, SFXTs exhibit lower average absorbing column densities than other Sg XBs, indicative of lower average density winds \citep{kretschmar19}.  The donor star in one SFXT, IGR J11215-5952, appears to have a magnetically focused supergiant wind, and it is possible that outbursts in this SFXT are associated with magnetic reconnection events between the neutron star magnetosphere and the magnetic field of the wind.  \par
An alternative, or perhaps complementary, explanation for the transient behavior of SFXTs is that it is associated with transitions between different accretion regimes, as described in \S\hyperref[sec:magnetosphere]{2.1.1}.  Magnetic or centrifugal gates to accretion may be responsible for the low luminosity states ($L_X\lesssim10^{33}$ erg s$^{-1}$) of SFXTs, while temporary increases in the accretion rate (which may be associated with wind clumps) can induce transitions to more efficient accretion regimes such as the subsonic settling accretion or direct Bondi-Hoyle accretion (\cite{bozzo08,kretschmar19}).  The vast majority of SFXT outbursts can be explained by transitions between different wind accretion regimes, with the exception of IGR J17544-2619, which has exhibited an extremely bright outburst reaching $L_X\sim3\times10^{38}$ erg/s which may require the formation of a temporary accretion disk (\cite{kretschmar19}, and references therein).  The ``off-states" of HMXBs including Vela X-1 and 4U 1907+09, during which the source flux drops below the instrumental sensitivity, may also result from magnetospheric gating, and thus provide a link between classical HMXBs and SFXTs (e.g. \cite{kreykenbohm08, doroshenko12}).

\subsection{3.3 Be X-ray Binaries}
\label{sec:bexrb}
\begin{figure}[t]
\centering
\includegraphics[width=0.3\textwidth, angle=90]{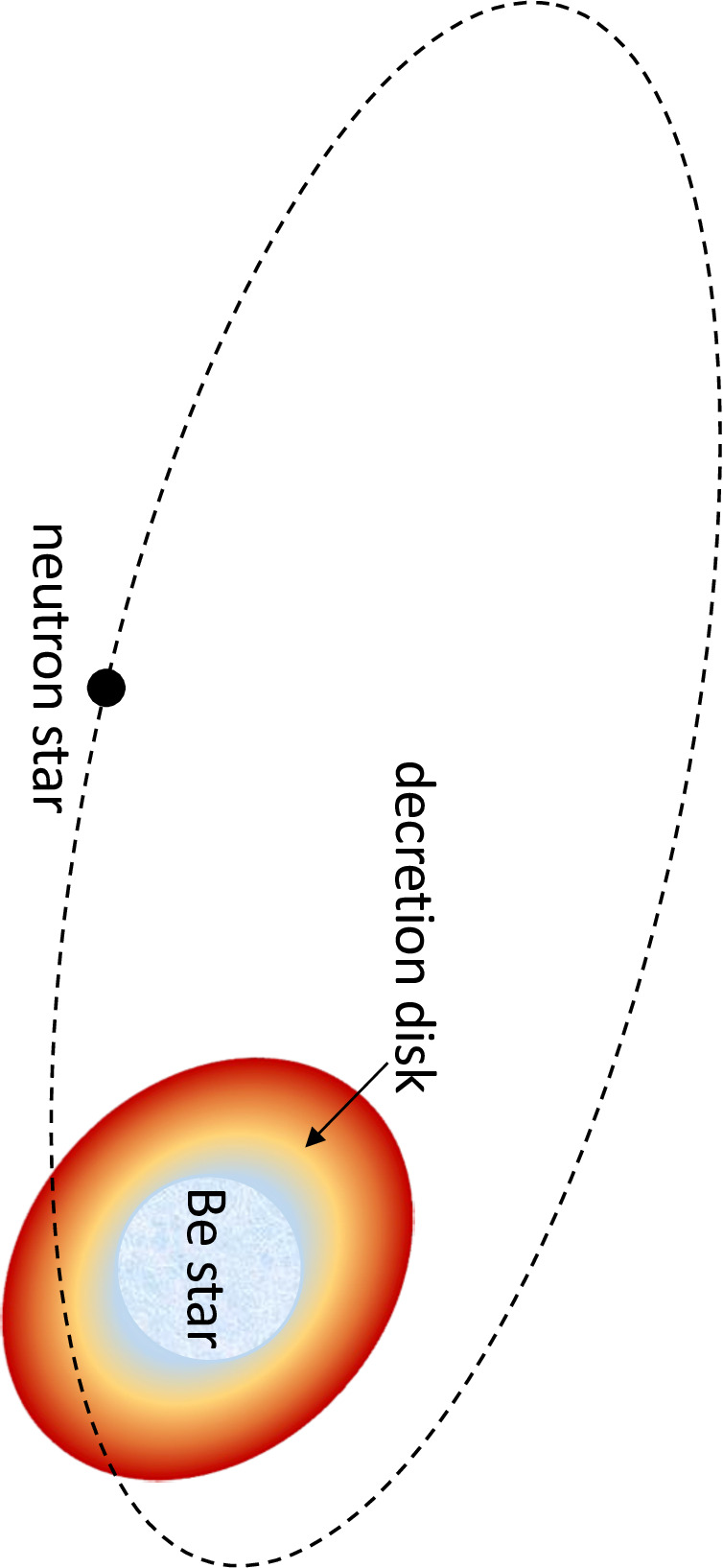}
\caption{Schematic diagram of a typical Be XB, consisting of a neutron star in an eccentric orbit around a Be star with an equatorial, cool decretion disk.  Type I outbursts from Be XBs are observed near periastron, as matter from the decretion disk is accreted by the NS.}
\label{fig:behmxb}       
\end{figure}

Be X-ray binaries (Be XBs) consist of a neutron star\footnote{In a few cases the compact object has been identified as a black hole or a white dwarf.} orbiting an Oe or Be star of luminosity class III-V.  As shown in Fig. \ref{fig:behmxb}, these donor stars have slow, equatorial winds often referred to as decretion disks.  Be XBs tend to have longer orbital periods than wind-fed Sg XBs ($P_{\mathrm{orb}}\gtrsim20$ days) and span a wide range of NS spin periods ($P_{\mathrm{spin}}\sim1-1000$ seconds).
These systems were originally considered transient XRBs (i.e. whose flux varies by at least 2 orders of magnitude with respect to their quiescent state), though recently several persistent systems (with \lx $\sim10^{33-35}$ \lxunit) have also been identified in our Galaxy and the Magellanic Clouds. For a review of these systems refer to \cite{reig11}, and references therein. \par
\begin{figure}[b]
\centering
\includegraphics[width=0.2\textwidth, angle=90]{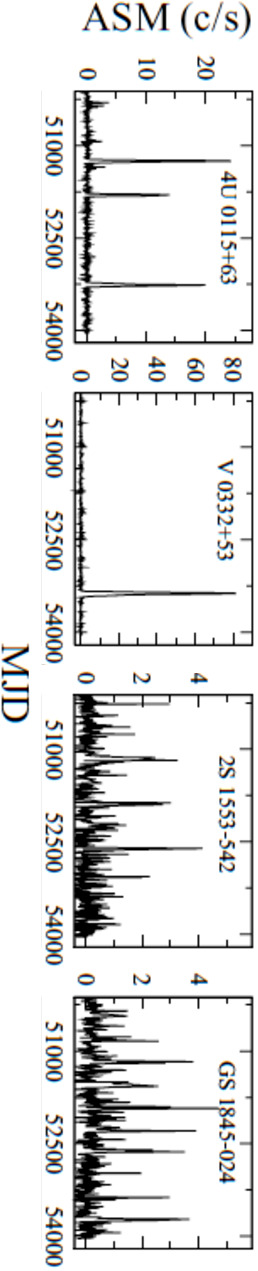}
\caption{\textit{RXTE} ASM light curves of some Be XBs from 1996 February to 2006 November, binned by 5 days. Type II outbursts can be observed in 4U 0115+63 and V 0332+53, while periodic Type I outbursts can be seen in 2S 1553-542 and GS 1845-024.  Credit: Taken from Figure 1 in Reig, P., ``On the neutron star-disk interaction in Be/X-ray binaries", MNRAS, 377, 2  \cite{reig07}.}
\label{fig:reig07}       
\end{figure}

\begin{figure}[t]
\centering
\includegraphics[width=0.6\textwidth]{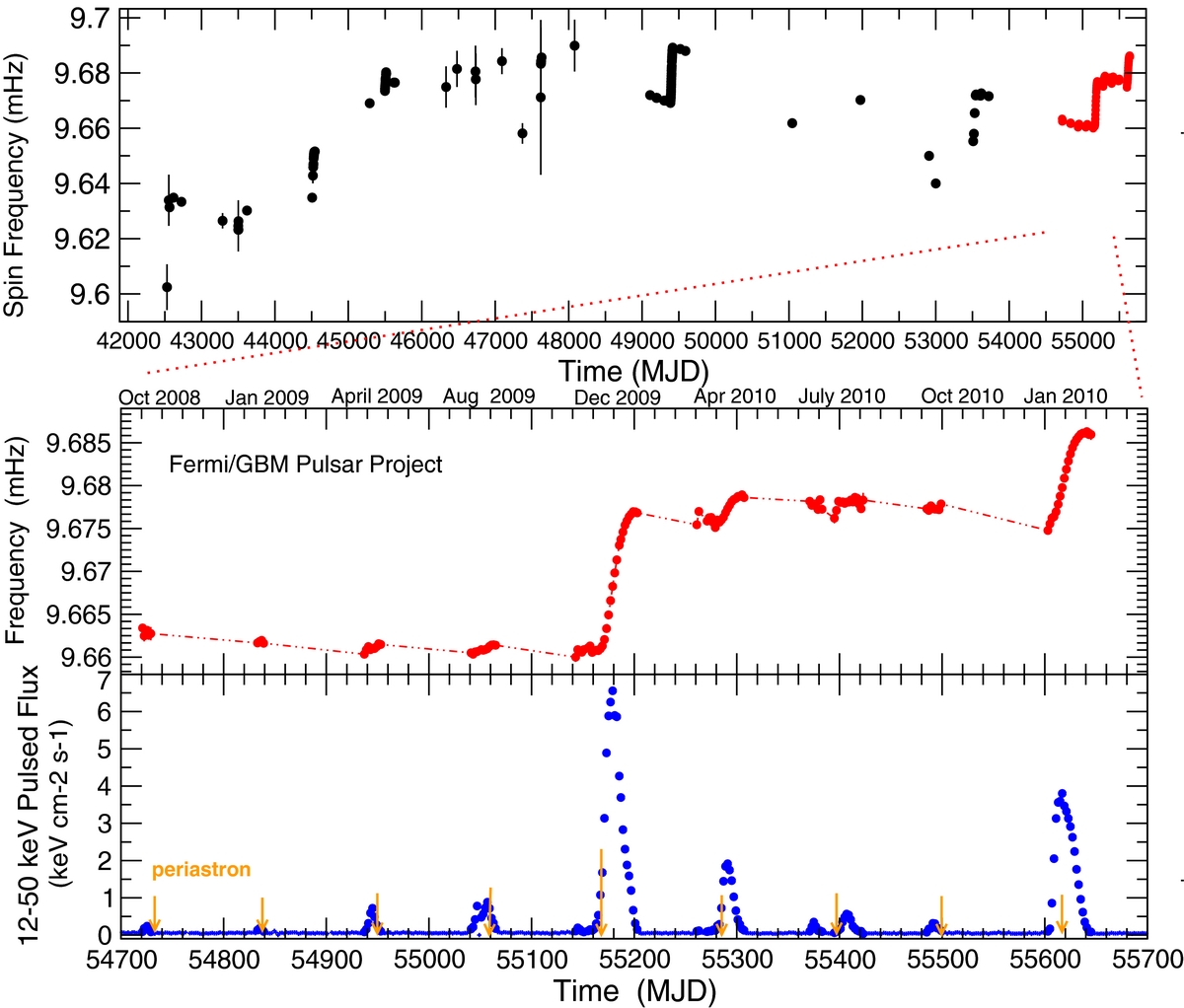}
\caption{\textit{Top:} Long-term pulsar spin frequency history of A 0535+26 since 1975. \textit{Middle:} Zoom-in of the frequency history since 2008 by \textit{Fermi} GBM. Frequency measurements are connected with a dot-dashed line. \textit{Bottom:} Daily average 12–50 keV pulsed flux measured with \textit{Fermi} GBM.  Periodic type I outbursts at periastron are observed, as well as a couple of bright Type II outbursts (also near periastron for this source) which coincide with sharp increases of the spin frequency as the NS is spun-up. Credit: Figure 5 in \cite{camero12}, reproduced with permission \textcopyright AAS.}
\label{fig:camero12}       
\end{figure}

The persistent Be XBs are usually pulsars with long spin periods (\pspin $\gtrsim$ 200 s) in wide (\porb $\gtrsim$ 200 days), eccentric ($e \lesssim 0.2$) orbits. In contrast, the transient Be XBs exhibit outbursting behavior, classified as either Type I or Type II outbursts.  Example lightcurves of some transient Be XBs are shown in Fig. \ref{fig:reig07}.  Type I outbursts happen (quasi-)periodically at or near periastron, are usually short lived, lasting only for a small fraction of the orbit, and increase the X-ray luminosity by about an order of magnitude.  Rapid pulsar spin-up rates are observed during some Type I outbursts, pointing to the creation of an accretion disk in these cases \cite{ziolkowski02}.  These outbursts have been explained as mass transfer from a Be star disk that has been tidally truncated.  In contrast, during Type II outbursts, the X-ray luminosity increases by a factor of $10^3-10^4$ above quiescence, sometimes reaching the Eddington luminosity for a neutron star ($L_X\sim10^{38}-10^{39}$ erg s$^{-1}$). These major outburst events are not modulated on the orbital period, can last for a long time (sometimes as long as several orbits), and can even lead to the disappearance of the Be circumstellar disk (as evidenced by measurements of the H$\alpha$ equivalent width in some of these systems). Other observational characteristics of Type II outbursts include the formation of an accretion disk around the neutron star, and in turn, large pulsar spin-up rates (that increase with increasing luminosity), as shown in Fig. \ref{fig:camero12}; in some systems quasi-periodic oscillations have also been discovered. These type of outbursts have been associated with unstable disk configurations such as warped disks \citep{haberl16}, and are more likely to be associated with rapidly spinning neutron stars or systems with low eccentricity and short orbital periods \citep{reig07}.

The X-ray properties of Be XBs, including their outburst behavior, is strongly tied to the properties of Be stars, so a detailed understanding of Be stars is critical to a complete view of Be XBs.  Be stars are fast rotators with reported values close or above about 75\% of the critical velocity that would result in the break-up of the star due to the centrifugal force at the surface matching the gravitational force \citep{rivinius13}.  It is debated whether these rotational velocities reflect their angular momentum distribution at birth or if the Be stars have been spun up as a result of binary evolution and mass transfer. Although the exact mechanism that triggers the gradual drift of matter outward from the Be star's equatorial region is not well understood, the formation of a decretion disk around the Be star (in which matter flows outward away from the Be star) is explained by the viscous decretion disk model \citep{lee91}, which offers insight into the Be star disks and their dynamical evolution.

Two of the main observational characteristics of Be stars in the optical and infrared bands have their origin in the circumstellar disk. These features are their infrared excess emission and their emission spectral lines (mainly the Balmer and Paschen series of hydrogen but sometimes also lines of He and Fe), with the most prominent line being H$\alpha$. 
Be stars appear redder than non-emitting B stars of the same spectral type due to additional circumstellar reddening caused by the hydrogen free-free and free-bound emission of the disk \citep{rivinius13}. 
The presence of the disk is useful in the identification of new Be stars.  For example, the presence of warm dust in the circumstellar envelope of the disk produces an IR excess at mid- and far-infrared wavelengths that can be used to identify new Be stars, as demonstrated by studies of high-mass stars in the LMC and SMC with \spitzer (e.g. \cite{bonanos09}).   During a disk-loss episode, the H$\alpha$ line is in absorption, and the X-ray activity is minimal. Thus, \ha\, surveys have been proven to be another effective way to identify Oe and/or Be stars. In the SMC, the fraction of Be XBs among their parent OBe population has been estimated to be $\sim$0.002-0.025 by means of broad-band R and narrow-band \ha\, photometry of the stellar populations \citep{maravelias17}. \par

The H$\alpha$ line is considered the best indicator of the decretion disk state. Studies of the line variability offer insight on the geometry (size, shape) and dynamics of the circumstellar envelope (\cite{reig11}, and references therein). Long-term \ha\, monitoring of 5 Galactic Be XBs showed that \textit{(i)} Type I outbursts generally occur when the Be disk radius is truncated at radii close to/larger than the Roche lobe radius at periastron passage (i.e. when the disk radius approaches the \textit{L1} Lagrangian point, where mass transfer can occur), while \textit {(ii)} Type II outbursts were not found to be either correlated or anti-correlated with the disk size \citep{monageng17}. The \ha\, line can have either a symmetric profile (pointing to quasi-Keplerian disks) or an asymmetric one (caused by distortions of the quasi-Keplerian disk, associated with one-armed global disk oscillations), and it can change from one profile type to the other on timescales of months to years. Most Be XBs display asymmetric split \ha\, profiles; the V/R ratio, which measures the violet- to red-side line peak intensities above the continuum, is often used to describe the line's asymmetry (e.g., \cite{reig11}).\par

\ha\ studies of Be XBs have shown that the Be circumstellar disk is affected by the presence of the neutron star. Using the equivalent width of the \ha\, line as a proxy of the size of the Be star's disk, it has been shown that the Be disk is truncated by the tidal/resonant interaction with the neutron star (e.g., \cite{coe15}).  Systems with short orbital periods have smaller disks as these are the systems that experience the tidal truncation more frequently and to a larger degree. In these cases, the disk density increases at a higher rate due to the resonant torque preventing the circumstellar material from drifting outwards.

\subsection{3.4 Wolf-Rayet X-ray binaries}

Although only a few HMXBs with Wolf-Rayet donor stars have been discovered, they are thought to represent an important evolutionary stage in the formation of double compact object systems and are thus at times treated as a distinct HMXB class.  Wolf-Rayet stars display strong, broad emission lines of ionized helium, nitrogen, carbon, or oxygen; they are evolved high-mass stars that have lost their outer hydrogen envelope and possess powerful, high-velocity winds.  Wolf-Rayet HMXBs are likely to be produced by the evolution of Sg XBs; when the donor star in an Sg XB evolves and expands, it is expected to result in a common envelope phase if the binary mass ratio is low \citep{vandenheuvel17}.  If the compact object is a NS, it is likely to spiral inwards and result in a merger, producing a Thorne-Zytkow object.  Instead, if the compact object is a BH, it can eject the common envelope and survive spiral in, resulting in a Wolf-Rayet BH binary with a short orbital period of less than a day. \par
Two Wolf-Rayet HMXBs has been identified in our Galaxy \cite{fortin23}: 
Cyg X-3 and OAO 1657-415.  The Wolf-Rayet nature of the donor star in Cyg X-3 was spectroscopically confirmed, and while the mass estimate of the compact object is consistent with either a neutron star or a low-mass black hole, the fact that it is a microquasar, displaying relativistic jets and changes in spectral state typical of accreting black holes (see \S\hyperref[sec:bhstates]{5.2}), favors the black hole possibility (\cite{zdziarski13} and references therein).  The donor star of OAO 1657-415 is classified as a Ofpe/WN9A supergiant .  In contrast to Cyg X-3, this HMXB hosts a pulsar; its accretion mode appears to switch between disk-fed and wind-fed accretion \citep{walter15}.  A few additional examples of Wolf-Rayet X-ray binaries have been discovered in other galaxies, including two ultraluminous X-ray binaries (see \S\hyperref[sec:ulx]{3.5}, e.g. \cite{liu13,esposito15}).  These HMXBs may evolve into binary BHs in close orbits, which may eventually merge and produce gravitational waves \citep{vandenheuvel17}.  

\subsection{3.5 Ultraluminous X-ray Sources}
\label{sec:ulx}
\hbindex{Ultraluminous X-ray sources} (ULXs) are point-like sources with X-ray luminosities exceeding $L_{\mathrm{X}}=1-2\times10^{39}$ erg s$^{-1}$ but which are not active galactic nuclei (AGN) (see reviews \cite{kaaret17, fabrika21}).  This X-ray luminosity threshold corresponds to the Eddington limit for a black hole with a mass of $10-15 M_{\odot}$, which are the highest dynamically-measured masses of black holes in HMXBs (see \S\hyperref[sec:masses]{4}).  The first ULXs were discovered with the \textit{Einstein Observatory}, and currently several hundred ULXs are known \citep{fabrika21}.  ULXs are primarily found in spiral or irregular galaxies, and appear to be associated with very young stellar populations and low-metallicity environments (\cite{kaaret17, fabrika21}). \par
Historically, two models for ULXs have been proposed: intermediate mass black holes ($M\sim10^3-10^5 M_{\odot}$) accreting at sub-Eddington rates or stellar-mass black holes in Roche-lobe overflow binary systems accreting at super-Eddington rates \citep{fabrika21}.  Some ULXs show evidence for powerful outflows in their X-ray and optical spectra; such outflows are expected to blow off the surface of geometrically thick disks when supercritical accretion occurs, thus indicating that some ULXs are super-Eddington accretors.  Additional evidence for super-accretors in ULXs comes from studies of the optical bubble nebulae that have been detected extending for tens to hundreds of parsecs around many ULXs; their emission line excitation ratios indicate the presence of shock waves arising from relativistic winds or jets colliding with the surrounding medium and, in some cases, the presence of a powerful photo-ionizing extreme ultraviolet source, all of which are features consistent with super-Eddington accretion.  One object in our Galaxy observed to accrete in super-Eddington mode, SS 433, is a microquasar with relativistic jets whose bolometric luminosity is estimated to be $\sim10^{40}$ erg s$^{-1}$, making it a ULX candidate even though its observed X-ray luminosity is low ($L_X\sim10^{36}$ erg s$^{-1}$) due to the inner parts of its accretion disk being shielded by an optically thick wind \citep{fabrika04}; this source is thus often considered an example of a ULX hosting a super-accreting black hole.  Quasi-periodic oscillations (QPOs) have been detected in 5 ULXs \citep{fabrika21}, suggesting that these sources also host stellar-mass black holes.  
 \par
The surprising discovery of coherent X-ray pulsations from M82 X-2 and a handful of other ULXs reveals that some ULXs host neutron stars accreting at rates a factor of $\sim10-100$ above the Eddington rate.  A few additional sources have been observed to be transient ULX pulsars, which may be neutron stars in eccentric orbits accreting from Be donors and experiencing super-Eddington accretion only during outbursts \citep{king19}.  The first Galactic ULX pulsar, Swift J0243.6+6124, was recently discovered during a super-Eddington outburst in 2017-2018 \citep{mushtukov22}.  An additional ULX, M51 X-8, is also thought to host a neutron star based on the detection of a cyclotron line in its X-ray spectrum \citep{fabrika21}.  It remains unclear what fraction of ULXs host neutron stars.  Since evolutionary models suggest that binary systems with neutron stars are $10-50$ times more numerous than those with black holes \citep{belczynski09}, some have argued that the majority of ULXs may host super-accreting neutron stars \citep{fabrika21}.  However, if super-Eddington accretion onto neutron stars requires special conditions, such as the presence of a very strong magnetic field, then the fraction of neutron stars among ULXs may be much lower.  \par
Taken together, these various pieces of evidence support the view that most ULXs are super-accreting black holes or neutron stars.  However, a small fraction of ULXs may host intermediate-mass black holes (IMBHs), especially some of the most luminous ULXs ($L_X>5\times10^{40}$ erg s$^{-1}$) whose X-ray spectra and variability more closely resemble those of BHs accreting at sub-Eddington rates.  One of the most compelling IMBH candidates discovered to date is ESO 243-49 HLX-1, a ULX classified as a hyperluminous X-ray source (HLX) due to its extreme luminosity, which can reach $1.1\times10^{42}$ erg s$^{-1}$.  \par
Combined with evidence indicating that most ULXs host stellar-mass black holes or neutron stars, studies of the stellar populations in the vicinity of ULXs have led to the consensus view that most ULXs can be considered the high-luminosity end of the HMXB population.  ULXs are found primarily in star-forming galaxies.  In starburst galaxies, they tend to be located near star clusters with ages of $\sim6$ Myr, and in galaxies with moderate star formation rates, ULXs tend to be found near OB associations with ages of $10-20$ Myr \citep{kaaret17}.  The donor stars in ULXs are thus likely to be high-mass stars.  Although most ULXs are located in such crowded fields that unambiguous counterpart identification is not possible, in some cases, it has been possible to identify and study the individual optical counterparts of ULXs.  For about 20 ULXs, the SEDs of their optical counterparts have been well-sampled.  The optical SEDs of bright optical counterparts are very blue, consistent with the SEDs expected for the winds of super-Eddington accretion disks, whose optical emission dominates over that of the donor star \citep{fabrika15}.  Fainter optical counterparts instead have SEDs consistent with those of supergiant A-G stars, so in these ULXs the donor star may be able to outshine the hot accretion disk wind due to lower accretion rates \citep{fabrika21}. The optical spectra of only about 10 ULXs have been well-studied to date; they exhibit broad emission lines with FWHM$\approx300-1600$ km s$^{-1}$ indicative of strong disk winds \citep{fabrika15}.  The donor stars of 9 ULXs have been spectroscopically classified: M101 ULX-1 has a Wolf-Rayet donor, ULXP NGC 7793 P13 has a B9 Ia supergiant donor, the ULX candidate SS 433 has an A 3-7 I supergiant donor, and 6 other ULXs appear to be red supergiants although 5 of these require additional confirmation \citep{fabrika21}.  \par
\begin{figure}[t]
\centering
\includegraphics[width=0.28\textwidth, angle=90]{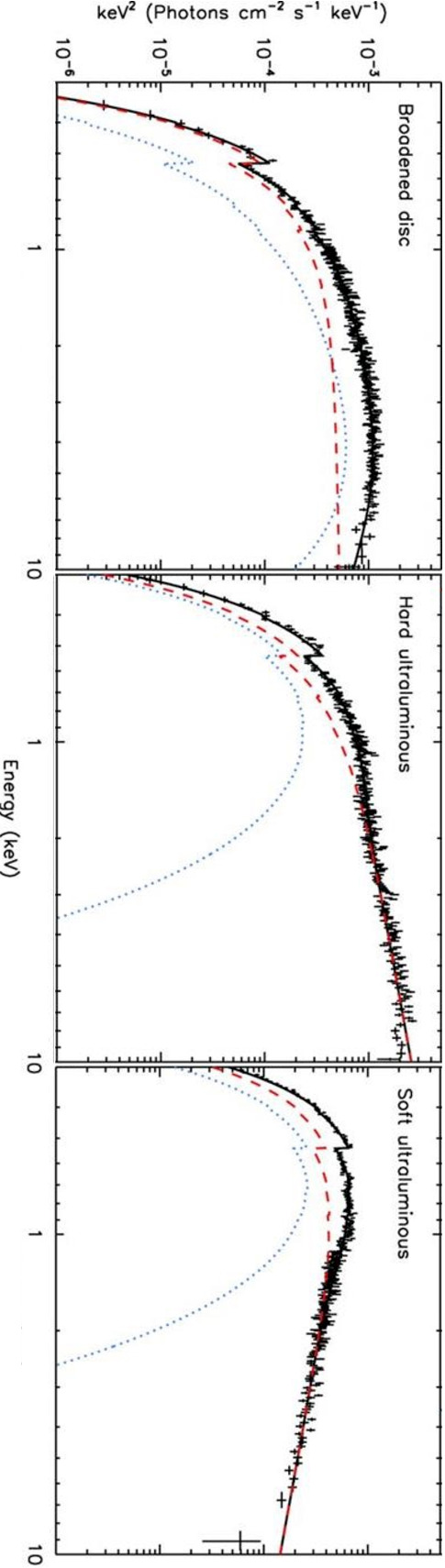}
\caption{Example energy X-ray spectra from \textit{XMM} EPIN PN observations of different ULXs exhibiting three different spectral states. From left to right they are: broadened disk – NGC 1313 X-2; hard ultraluminous - Ho IX X-1; soft ultraluminous - NGC 5408 X-1. The contributions from each of the components of the best-fitting absorbed multicolor disk (blue dotted line) plus power-law (red dashed line) model are shown. Credit: Figure 1 in \cite{sutton13}, ``The ultraluminous state revisited: fractional variability and spectral shape as diagnostics of super-Eddington accretion", MNRAS, 435, 2 }
\label{fig:sutton13}       
\end{figure}
Even though most ULXs are thought to be high-luminosity HMXBs, their X-ray spectral properties differ in significant ways from those of most HMXBs (see \S\hyperref[sec:spectra]{5.1}) due to having super-Eddington accretion disks.  ULX X-ray spectra often show a double-peaked shape and an exponential cutoff above 10 keV \citep{kaaret17}; the low energy peak is associated with the thermal temperature of the disk ($kT_{\mathrm{disk}}\sim0.2$ keV) while the second is due to the Comptonization of soft photons by free electrons with $kT_e\sim1-2$ keV and high optical depth ($\tau\geq6$).  For comparison, typical values for black hole HMXBs are $kT_{\mathrm{disk}}\sim1$ keV, $kT_e\sim100$ keV, and $\tau\leq1$ \citep{mcclintock06}, although it should be noted that it is not currently possible to constrain the very hard X-ray spectrum of ULXs well due to their low observed fluxes compared to Galactic HMXBs.  \par
Depending on the relative contributions of the different spectral components, ULX spectra can appear softer or harder, and ULXs can transition between different spectral states, which are shown in Fig. \ref{fig:sutton13}.  These different spectral states may result from the observer's line-of-sight relative to the central conical funnel formed by the disk wind; this angle can vary with the precession of the disk, thought to occur on timescales of $2-200$ days based on the detection of superorbital periods, and the accretion rate, which affects the disk geometry (\cite{kaaret17, fabrika21}).  The greatest variability is observed when ULXs are in the soft spectral state, but the majority of the variability is associated with the hard component \citep{fabrika21}.  This variability is consistent with the soft state being associated with a greater inclination between the line-of-sight and the wind funnel axis, which results in the inner accretion disk being more likely to occasionally be obscured by optically thick clumps of wind. 

\subsection{3.6 Gamma-ray binaries}
\label{sec:gammaraybinaries}
While X-ray emission generally dominates the spectral energy distribution of HXMBs, \hbindex{ gamma-ray binaries} radiate more power in gamma rays above 1 MeV than in X-rays. Such systems can be inconspicuous in X-rays yet stand out in the gamma-ray sky. 
Emission at such high energies is necessarily from non-thermal particles: in these systems, particle acceleration dominates how energy is channelled into radiation.  Like HMXBs, gamma-ray binaries consist of a compact object and a high-mass stellar companion, but a significant difference between these two classes of sources is that some (or all) gamma-ray binaries may be rotation-powered rather than accretion powered.  While differing in terms of energy production mechanisms, there may be an evolutionary link between gamma-ray binaries and HMXBs. 
 Table\,\ref{tab:binaries} lists confirmed gamma-ray binaries as of early 2022 (see \cite{dubus13,paredes19a,chernyakova20} for reviews and references).

\begin{table*}[b]
\centering
\begin{tabular}{@{}lccr@{}}
\toprule 
name & \multicolumn{2}{c}{binary components}  & P$_{\rm orb}$ (d)  \\
 \toprule 
PSR J2032+4127	& pulsar	& B0Ve & $\sim 17000$ \\
PSR B1259-63 	 	& pulsar 	& O9.5Ve & 1237 	\\ 
HESS J0632+057	& ? 		& B0Vpe & 315  	\\
LS I +61$^{\circ}$303& pulsar ? 		& B0Ve & 26.5 	\\
1FGL J1018.6-5856	& ? 		& O6V(f) & 16.6 	\\
4FGL J1405.1-6119	& ?		& O6.5III & 13.7	\\
LMC P3			& ?		& O5III(f) & 10.3	\\
LS 5039 		 	& pulsar ? 		& O6.5V(f) & 3.9	\\
\midrule 
\end{tabular}
\caption{Confirmed gamma-ray binaries\label{tab:binaries} (\cite{dubus13,paredes19a,chernyakova20}).}
\end{table*}

\begin{figure}[t]
\centering
\includegraphics[height=6cm]{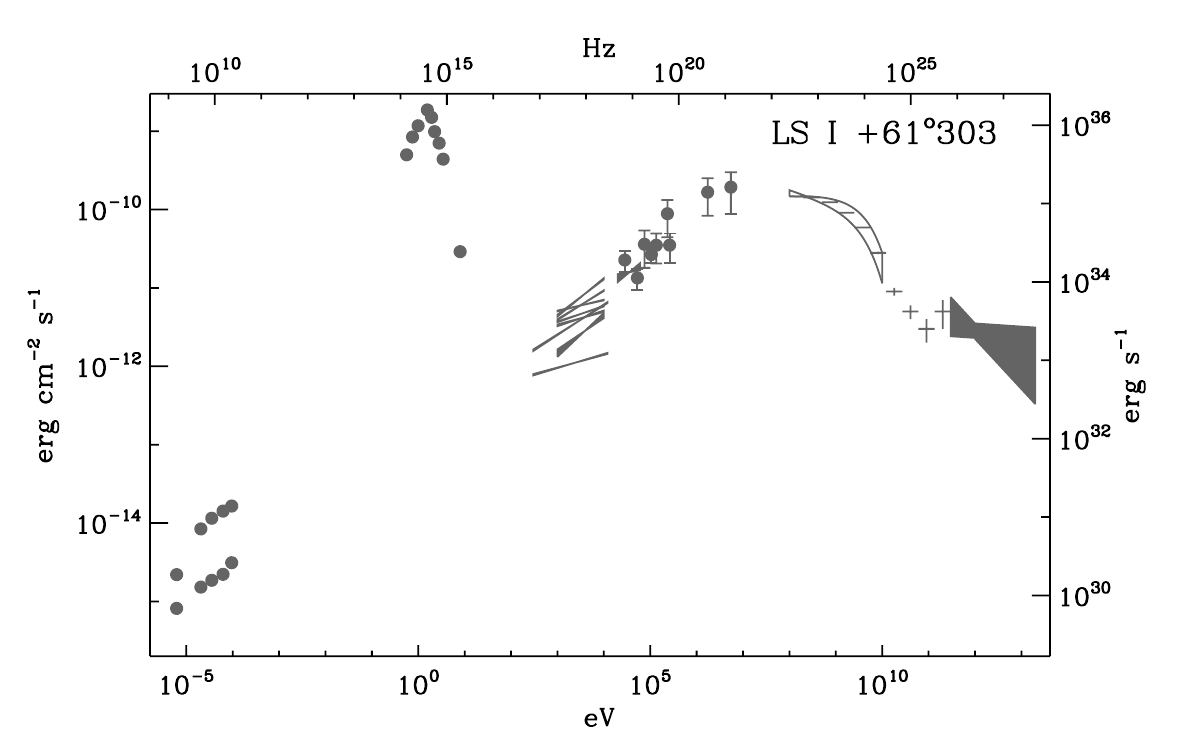}
%
%
\caption{Spectral energy distribution $\nu F_{\nu}$ of the gamma-ray binary LSI +61$^\circ$303 showing emission from radio to TeV gamma rays, peaking around 10 MeV (data from \cite{dubus13,veritas16,chernyakova17,xing17}).}
\label{fig:sed}       
\end{figure}

Gamma-ray binaries emit from radio to TeV energies, as shown in Fig.\,\ref{fig:sed}. Their radio emission is generally optically thin with an intensity $S_{\nu}\sim \nu^{-1/2}$ and is observed on a scale larger than the binary orbit. VLBI observations have resolved the radio emission of several binaries into a collimated flow whose position angle rotates with orbital phase, as expected if the non-thermal particles were collimated by the stellar wind of the companion. The optical-UV is dominated by radiation from the massive star. In X-rays, gamma-ray binaries typically show hard power-law spectra with photon indices $\Gamma\approx1.5$ extending well beyond 10 keV, whereas typical X-ray binaries show cutoffs (see \S\hyperref[sec:spectra]{5.1}). The spectral energy distribution of gamma-ray binaries peaks in the 0.1 to 10\,MeV range. 
In the higher 0.1 to 30\,GeV range, the {\em Fermi}-Large Area Telescope (LAT) spectra of these sources are usually well-fit by a power-law with an exponential cutoff at a few GeV. Above 100 GeV, the typical spectrum is a power-law with a photon index of $\Gamma\approx-2.5$.  Several particle acceleration sites and/or radiative processes must be involved given the clearly distinct spectral components from X-ray to TeV energies.

Two of the gamma-ray binaries, PSR B1259-63 and PSR J2032+4127, host radio pulsars with pulse periods of 48 ms and 143 ms respectively, a much faster rotation rate than that of HMXB X-ray pulsars (see \S\hyperref[sec:pulsations]{6.3.1}). Radio pulsar timing indicates that the neutron star in these two systems is steadily spinning down on timescales of $10^5$\,years, consistent with `normal' radio pulsars. Normal radio pulsars are {\em rotation-powered}: the combination of their fast rotation and high magnetic field ($\sim 10^{12}$\,G) generates a tenuous relativistic wind that extracts the rotational energy of the neutron star and accelerates particles to very high energies. Part of this spindown power is radiated in pulsed radio to gamma-ray emission in the vicinity of the neutron star, or further away in a pulsar wind nebula (PWN) where the relativistic wind interacts with the interstellar medium. 
The general understanding is that PSR B1259-63 and PSR J2032+4127 are scaled-down PWN, with the relativistic wind interacting with the companion's stellar wind on AU scales rather than the parsec scales of typical PWN. 

In most HMXBs, the neutron star accretes material from the stellar wind forming X-ray pulsations as the material is channelled towards the magnetic poles of the neutron star. The mechanism responsible for radio pulsations is not expected to occur in HMXB X-ray pulsars because electric fields are shorted out in this high density environment. Inversely, accretion is quenched if the pressure from a rotation-powered pulsar relativistic wind exceeds the ram pressure of the stellar wind at the Bondi-Hoyle-Littleton capture radius. This occurs when 
\begin{equation}
\dot{E}>4\dot{M} v_{\rm w} c\approx 10^{35}\,  \left(\frac{\dot{M}}{10^{16}\,\mathrm{g\,s}^{-1}}\right)\left(\frac{{v}_{\rm w}}{1000\,\mathrm{km}\,\mathrm{s}^{-1}}\right)\,\mathrm{erg\,s}^{-1}
\end{equation}
where $\dot{M}$ is the accretion rate and $v_{\rm w}$ is the stellar wind velocity. In PSR B1259-63 and PSR J2032+4127, the pulsar relativistic wind is powerful enough to quench the accretion flow and turn on the radio pulsar mechanism. As the pulsar spins down, its $\dot{E}$ decreases and its wind cannot hold off accretion. Thus, these pulsars are expected to switch from rotation-powered radio pulsars to accretion-powered X-ray pulsars on their spindown timescale of $\approx 10^5$\,yr.  

 All the gamma-ray binaries could be rotation-powered by young pulsars, like PSR B1259-63 and PSR J2032+4127. In this scenario, gamma-ray binaries represent a short-lived ($10^5$ yr) phase between the supernova birth of a fast-rotating neutron star and the longer-lived HMXB phase triggered by the onset of accretion. This is supported by the scarcity of gamma-ray binaries compared to HMXBs and by observational similarities between all gamma-ray binaries \citep{dubus17}. Radio pulsations may be very difficult or impossible to detect in most gamma-ray binaries: the dense environment due to the stellar wind or circumstellar disk scatters and attenuates radio pulsations, as happens for PSR B1259-63 over part of its eccentric orbit. Indeed, PSR B1259-63 and PSR J2032+4127 have the longest orbits amongst gamma-ray binaries, corresponding to the widest separations with their high-mass stellar companion. Pulsations can also be searched for at X-ray or gamma-ray energies, but the low photon count rates require integrating over tens to hundreds of orbits to build up sufficient signal-to-noise, so the pulsar orbital motion must be taken into account in the analysis.  Despite these hurdles, intermittent 269 ms radio pulsations have been attributed to LS I+61$^{\circ}$303 \citep{weng22}, and a tentative 9s gamma-ray pulsation has been reported for LS 5039 \citep{yoneda20}.

Alternatively, gamma-ray binaries have also been interpreted as microquasars, with the non-thermal emission arising from particle acceleration in a relativistic jet, in a scaled-down version of the high-energy emission observed from AGN jets (\cite{boschramon09,massi20}). The microquasar scenario would be supported if the compact object mass in some gamma-ray binaries turned out $\geq 3$\,M$_\odot$, pointing to a BH rather than a pulsar. In fact, three BH candidates with high-mass stellar companions are firmly detected at gamma-ray energies: Cyg X-1, Cyg X-3, and SS 433, all of which are accreting and launch relativistic jets. However, the gamma-ray luminosities of Cyg X-1 and Cyg X-3 are only a few percent of their X-ray luminosities so, unlike gamma-ray binaries, these sources are not dominated by non-thermal emission. Their gamma-ray emission is also correlated with changes in their radio/X-ray spectral state, clearly linking particle acceleration to the accretion/ejection process in these sources.  Gamma-ray binaries do not show these spectral state changes typical of accreting BH X-ray binaries. In the specific case of SS 433, the gamma-ray emission region is not the binary itself but a region where the jet-inflated bubble interacts with the ISM. 

Gamma-ray surveys and radio pulsar surveys are the prime tools to discover new gamma-ray binaries. In gamma-rays, point sources in the Galactic Plane make good candidates, triggering follow-up observations in the radio, X-ray, and optical bands to search for a possible HMXB association. The angular resolution is not always sufficient to rule out a chance superposition with another type of source,
so variability is key to establish the association of the gamma-ray emission with the binary. Searching for periodic emission in unidentified gamma-ray sources of the {\em Fermi}-LAT catalog has proven fruitful to discover new gamma-ray binaries.
PSR B1259-63 and PSR J2032+4127 were instead discovered in radio pulsar searches, where they stood out because of their massive stellar companion. Their powerful gamma-ray emission was only detected later, identifying them as gamma-ray binaries. 
Future radio surveys with the SKA can be expected to increase the number of known pulsars with massive companions, providing a new path to identifying gamma-ray binary candidates.

\subsection{3.6 Black Hole versus Neutron Star X-ray binaries}
\label{sec:bhns}
There are several observational signatures that can help determine whether the compact object in an HMXB is a NS or a BH.  The detection of X-ray pulsations (see \S\hyperref[sec:pulsations]{6.1.3}) is considered definitive evidence of the presence of a NS, and such objects are referred to as pulsars.  These X-ray pulsations result from the X-ray emission produced near the star's magnetic poles coming into and out of the observer's view as the NS rotates.  Since the accretion flow around BHs is not collimated by magnetic fields, no X-ray pulsations are expected from BH HMXBs.  However, the lack of pulsations does not necessarily mean that an HMXB harbors a BH since a neutron star's magnetic field geometry, the degree of misalignment between its magnetic and rotation axes, the orientation of its magnetic poles relative to our line-of-sight, or a very long spin period can make X-ray pulsations difficult to detect.   \par
X-ray spectral features of HMXBs, described in \S\hyperref[sec:spectra]{5.1}, can also help to differentiate between NSs and BHs (see \S\hyperref[sec:spectra]{5.1}), although many of these diagnostics require fairly high quality X-ray spectra.  Some NS HMXBs exhibit cyclotron resonance scattering features in their X-ray spectra, which are produced as a result of their strong magnetic fields and thus provide clear evidence of the presence of a NS.  NS HMXBs tend to exhibit exponential cutoffs to their power-law spectra at lower energies than BH HMXBs, so the detection of an exponential cutoff at $\lesssim40$ keV is indicative, but not definitive, evidence of a NS.  Furthermore, accreting BHs undergo spectral state transitions, while most NS HMXBs do not.  In the high/soft state, the spectra of BH XBs exhibit relativistic reflection features from the inner edge of the accretion disk, including a relativistically broadened iron line whose width depends on the BH spin.  In the low/hard state, BH XBs exhibit a tight correlation between their X-ray and radio emission \citep{gallo03}, the latter of which is associated with relativistic jets.  Thus, observing that an HMXB undergoes spectral state transitions, possesses the aforementioned spectral features, or obeys the X-ray radio correlation provides a strong, but not definitive, indication of the presence of a BH. 

\subsubsection{3.6.1 Mass measurements of compact objects in HMXBs}
\label{sec:masses}
\begin{figure}[t]
\centering
\includegraphics[width=0.5\textwidth]{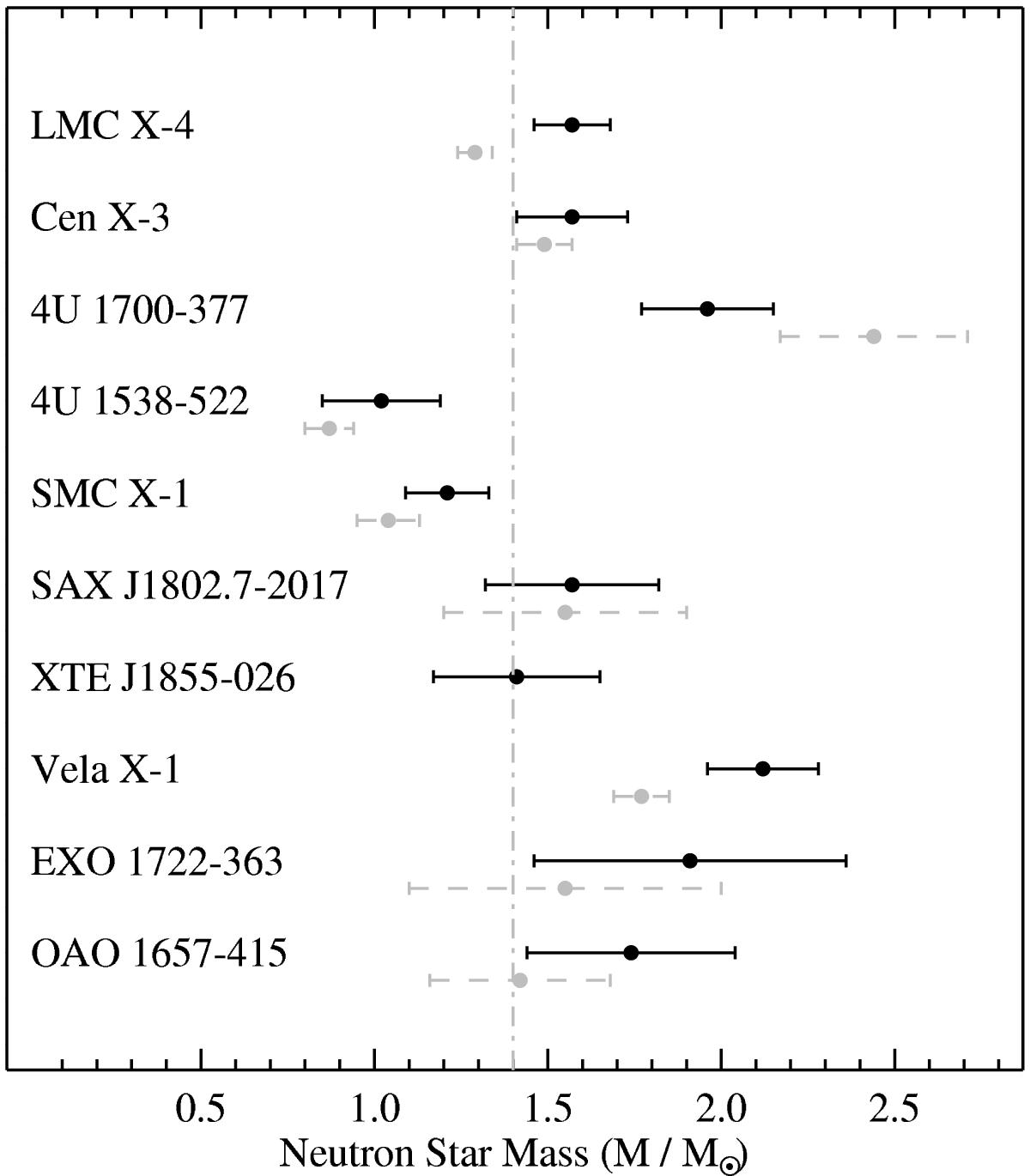}
\caption{Masses of the ten eclipsing HMXBs. The neutron star masses determined by \cite{falanga15} are shown with solid lines, while values from prior literature are represented with dashed lines. The error bars correspond to uncertainties at 1$\sigma$ confidence level. The dashed vertical line indicates the canonical neutron star mass of 1.4 M$_{\odot}$. Credit: Figure 4 from Falanga, M., et al., A\&A, 577, A130, 2015, reproduced with permission \textcopyright\ ESO.}
\label{fig:falanga15}       
\end{figure}

Measuring a compact object mass that exceeds $2-3$ \msun\ currently provides the most definitive observational proof of a BH since NS masses cannot theoretically exceed such values.  However, even this method is imperfect in that it cannot reliably differentiate between massive NSs and low-mass BHs.  The masses of compact objects in HMXBs can be constrained in two primary ways: (i) using the X-ray lightcurves of eclipsing binaries, which have a sufficiently high inclination that the compact object is eclipsed by its companion or (ii) measuring the radial velocity curve of the donor star.  Using more than ten years of monitoring observations by the Integral Soft Gamma-Ray Imager (ISGRI) on the \textit{INTEGRAL} satellite and the All-Sky Monitor (ASM) on \textit{RXTE}, \cite{falanga15} measured the masses of ten eclipsing HMXBs.  The masses were constrained by determining the binary orbital parameters, and then measuring the duration of X-ray eclipses in the $17-40$ keV lightcurves, folded on the orbital period.  Most of the HMXB eclipsing sample consisted of supergiant HMXBs with orbital periods $\leq$10 days, since it easier to discover and accurately measure eclipses for short-period systems.  As shown in Fig. \ref{fig:falanga15}, the mass range of the compact objects in these HMXBs was found to be $1.02-2.12$ \msun\, with the majority of the masses clustering between values of $1.4-1.7$ \msun\, as expected for neutron stars.  \par
Radial velocity measurements of the donor star in an HMXB can provide a lower limit on the compact object mass, if the mass of the donor can be estimated based on its spectral type.  If the inclination angle of the binary can be constrained, then this lower limit on the compact object mass can be tightened.  X-ray polarization measurements, which can be made with unprecedented sensitivity with the Imaging X-ray Polarimetry Explorer (IXPE), can help determine the inclination of X-ray emitting regions and piece together the geometric orientations of the binary components.  If an accretion disk is present around the compact object and the radial velocity of emission lines originating from the disk can be measured, or if changes in the timing of X-ray pulsations can be measured, then the compact object mass can be measured.  \par
The black hole mass constraints determined from radial velocity measurements for six HMXBs were compiled by \cite{ozel10}.  The orbital periods of these HMXBs range from $1-5$ days, and the donor stars include two Wolf-Rayet stars, two OB supergiants, and two OB giants/subgiants.  More recently, the mass of the compact object in MWC 656 was measured to be $3.8-6.9$ \msun\ \citep{casares14} based on the radial velocity curves of the Be donor star and the accretion disk; this mass measurement confirmed this source as the only currently known Be-BH XB.  This binary, originally identified through its uncertain association with a gamma-ray source, was found to be in an X-ray quiescent state ($L_X\sim 10^{-8}~ {\rm L_{Edd}}$), and confirmed to follow the X-ray/radio correlation \citep{ribo17}. \par
The masses of the black holes in HMXBs tend to be higher than those of black holes in LMXBs \citep{ozel10}.  This trend may result from the fact that mass transfer in LMXBs requires close binary separations that can likely only be produced if the system underwent a common envelope phase when the primary expanded; the ejection of the hydrogen envelope during this phase results in a naked helium core, which can experience further substantial mass loss during the Wolf-Rayet phase.  Thus, such a star would likely only be able to leave behind a black hole with mass $<10$ \msun\, but such a limit is unlikely to apply to HMXBs, which do not necessarily need to undergo a common envelope phase prior to the formation of the compact object.

\subsubsection{3.6.2 On the ratio of NS to BH HMXBs}
Out of the $\sim300$ HMXBs that have been identified in the Milky Way and Magellanic Clouds, more than half have been found to be pulsars \citep{liu06, coe15, haberl16, fortin23, neumann23}.  Analysis of the \chandra X-ray Visionary Program (XVP) observations of the SMC revealed that actually all \chandra sources with ${\rm L_{X}} \gtrsim 4 \times 10^{35}$\ergs\, exhibited X-ray pulsations \citep{hong17}.  In contrast to this large number of confirmed NSs, only a handful of HMXBs have been confirmed to be BHs based on compact object mass measurements: Cyg X-1, LMC X-1, LMC X-3, and MWC 656 \citep{ozel10, casares14}.  In addition, Cyg X-3 is considered likely to host a low-mass BH based on its X-ray and radio properties \citep{zdziarski13}.  Furthermore, V4641 Sgr, which hosts a $\sim6$ \msun\ BH accreting via RLO from its giant B9 companion, may be added to this list, although this source is sometimes classified as an LMXB because the stripped donor star is measured to currently have a mass of $\sim3$ \msun\ \citep{macdonald14}.  \par
While a large fraction of HMXBs still lack a definitive NS or BH classification, the available evidence points to the NS/BH fraction among HMXBs being quite high.  Given that Be XBs are more numerous than Sg XBs, especially in the Magellanic Clouds, the fact that only one Be-BH binary, MWC 656, has been discovered thus far suggests that the NS/BH fraction among Be XBs is particularly high.  This high NS/BH fraction may be due to a combination of stellar evolution and observational bias against X-ray faint systems.  The most obvious factor is that the initial stellar mass function naturally produces more NSs than BHs since the formation of a BH requires a higher initial stellar mass.  Furthermore, population synthesis studies have found that the Be-NS progenitor binaries have a higher probability of surviving the common envelope phase than Be-BH progenitor systems, and that only $\sim1-30$ Be-BH binaries may reside in the Galaxy \citep{belczynski09, grudzinska15}.  In addition to Be-BH binaries being rarer than Be-NS, some studies suggest they may also be $\gtrsim10$ times fainter than Be-NS XBs as a result of effective truncation of the Be circumstellar disk (e.g. \cite{brown18}).  The X-ray quiescent state of MWC 656 appears consistent with these predictions.

\section{4 Emission Properties}
Regardless of whether they host accreting NSs or BHs, the spectral energy distributions (SEDs) of HMXBs are dominated by emission arising from accretion onto the compact object at X-ray energies and by the emission of the donor star at optical-UV wavelengths.  The optical-UV and X-ray peaks of the HMXB SEDs are comparable to one another; the optical-UV luminosities of O/B Sg and Be stars are $L\sim10^{36}-10^{38}$ erg s$^{-1}$, which is similar to the X-ray luminosities of all but the brightest HMXBs.  In contrast, the SEDs of LMXBs are dominated by the X-ray emission associated with the accretion process.  \par
Some HMXBs exhibit significant emission in other energy bands.  Be HMXBs display an excess of mid-far infrared emission associated with the circumstellar disk around the Be star.  About 20 Galactic HMXBs are known to produce significant radio emission and 8 produce gamma-ray emission; the radio and gamma-ray emission is associated with relativistic jets (see \S\ref{sec:bhns} and \S\ref{sec:gammaraybinaries}).  Here we describe the X-ray emission of HMXBs and how the spectra of NS and BH HMXBs differ.

\subsection{4.1 NS HMXB X-ray spectra}
\label{sec:spectra}
\begin{figure}[b]
\centering
\includegraphics[width=0.7\textwidth]{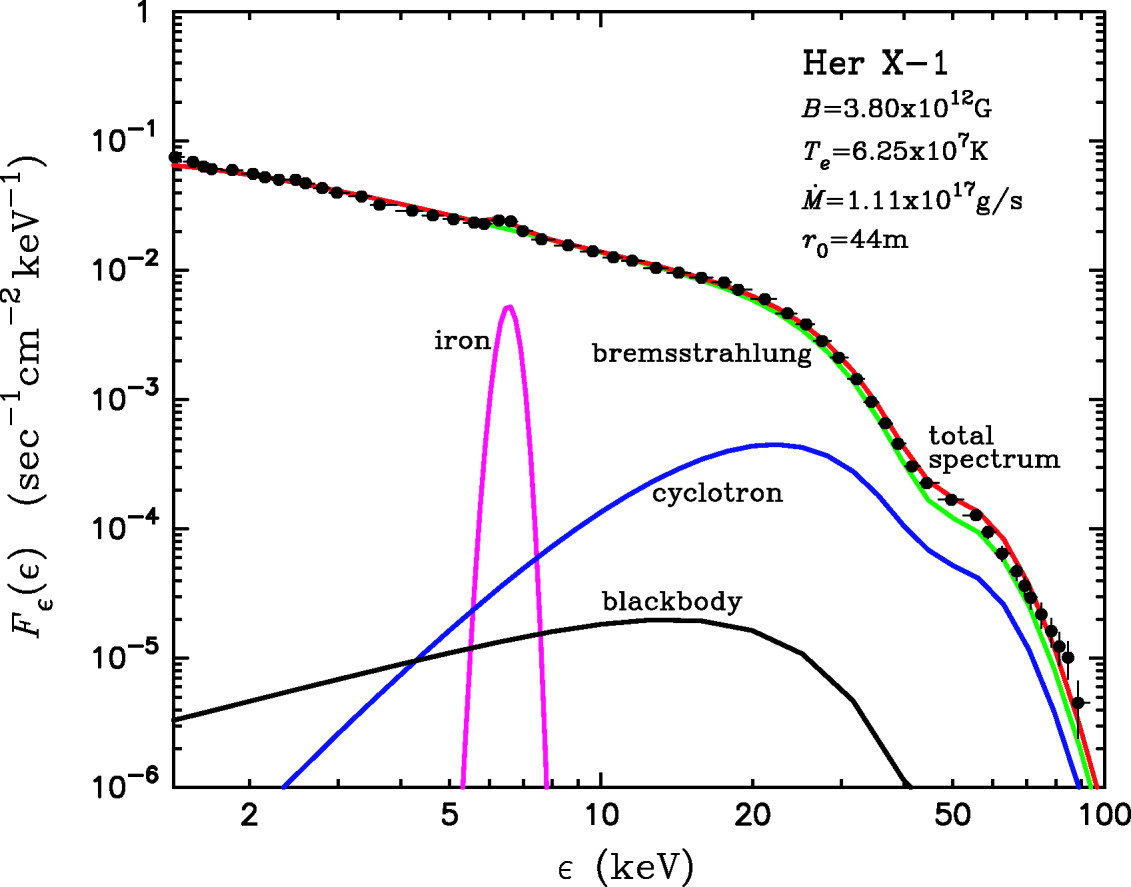}
\caption{Incident X-ray spectrum (deconvolved from \textit{BeppoSAX} effective area) for Her X-1, a NS HMXB (circles) compared to spectral model developed by \cite{becker07}.  Lines display the total modeled spectrum (red), as well as the individual components due to Comptonized bremsstrahlung radiation (green), Comptonized cyclotron emission (blue), Comptonized blackbody radiation (black), and the iron emission line (magenta).  Key best-fitting model parameters are reported in the top right corner: the magnetic field strength $B$, the temperature of the electrons in the optically thin region above the thermal mound located at the polar cap on the NS surface $T_e$, the accretion rate $\dot{M}$, and the accretion column radius $r_0$.
Credit: Figure 6 from \cite{becker07}, reproduced with permission \textcopyright AAS.}
\label{fig:becker07}       
\end{figure}
At X-ray energies, accreting neutron stars in HMXBs typically exhibit power-law spectra with photon indices of $\Gamma\sim0-2$ and exponential cutoffs with cutoff energies of approximately $10-20$ keV and e-folding energies of $\sim10$ keV \citep{coburn02}.  Since neutron stars in HMXBs have high magnetic fields ($B\sim10^{12}$ G), the accreting material is channeled to the magnetic poles, and the X-ray emission originates primarily in regions close to the magnetic poles. Electrons in the accretion flow, due to their thermal and bulk motion, Compton scatter seed photons to higher energies; the seed photons may be produced by a variety of radiative processes, including blackbody emission from the neutron star polar cap, and bremsstrahlung and cyclotron emission within the accretion column or accretion shocks \citep{becker07}.  The magnetic field strengths of accreting neutron stars can be measured from the cyclotron absorption lines that appear in some of their spectra; these features are discussed in more detail in \S\hyperref[sec:cyclotron]{5.1.1}.  Fig. \ref{fig:becker07} displays the \hbindex{X-ray spectrum} of a NS HMXB and the spectral contributions from these different emission processes.  Due to the complexity of the emission mechanisms in these systems, no self-consistent physical model has yet been developed that can successfully describe NS HMXB spectra across a broad range of mass accretion rates \citep{mushtukov22}. \par
At least one NS HMXB, 4U 2206+54, has been found to exhibit a hard power-law spectrum ($\Gamma\approx1-2$) extending up to $\sim150$ keV, with no evidence of a cutoff \citep{reig12}.  Such a spectrum resembles that of anomalous X-ray pulsars (AXPs) which are thought to be magnetars with $B\sim10^{13}-10^{15}$ G).  The possibility that 4U 2206+54 hosts a magnetar is further supported by its slow spin period ($P_{\mathrm{spin}}=5560$ s), and a few other HMXBs with slow pulsations ($P_{\mathrm{spin}}>500$ s) may host magnetars as well. \par
A soft X-ray excess is observed in some NS HMXB spectra, which is often modeled with a blackbody component \citep{mushtukov22}.  In disk-fed HMXBs, this soft excess likely results from the reprocessing of hard X-rays by the inner edge of the accretion disk.  In wind-fed HMXBs, the soft excess may arise from the photoionized stellar wind in the vicinity of the NS or thermal emission from the NS surface.
Both NS and BH HMXB spectra can display iron (Fe) fluorescent emission lines.  This fluorescent emission is produced when an X-ray photon ejects an inner shell electron, and then an electron from an upper energy level falls down to a lower energy level.  The most common Fe fluorescent line observed in HMXB spectra is Fe K$\alpha$, which is produced when an electron falls down to the lowest energy level ($n=1$) from the second energy level ($n=2$).  In NS HMXBs, Fe K$\alpha$ emission lines have typical equivalent widths of $\sim100$ eV and a central line energy of 6.4 keV, corresponding to neutral or low-ionization Fe \citep{coburn02}.  This iron emission can provide insights into the geometric structure of material around the neutron star \citep{mushtukov22}. 

\subsubsection{4.1.1 Cyclotron resonance scattering features}
\label{sec:cyclotron}

Some NS HMXB spectra exhibit cyclotron resonance scattering features (CRSFs) in their spectra \citep{coburn02}.  CRSFs are absorption line-like features arising from the resonant scattering of photons by electrons whose energies are quantized into Landau levels by a strong magnetic field.  The quantized energy levels are approximately harmonically spaced, with the fundamental line energy begin equal to:
\begin{equation}
E_c = 11.6 \frac{B}{10^{12} \mathrm{\,G}} (1+z)^{-1} \mathrm{\,keV} 
\end{equation}
where $B$ is the magnetic field strength and $z$ is the gravitational redshift of the NS.  Since the sizes and masses of neutron stars span a narrow range, the fundamental cyclotron line energy can be used to estimate the strength of the NS magnetic field.  One complication in the measurement of the fundamental line energy is that in hard sources with multiple strong harmonic lines, the fundamental line can be difficult to detect due to photon spawning, when an electron remains in an excited Landau level after scattering and emits a photon of similar energy to the fundamental line energy when it de-excites. \par
Using cyclotron line measurements of 10 accreting neutron stars, \cite{coburn02} found a positive correlation between the magnetic field strength and the power-law cutoff energy of the NS X-ray spectra up to CRSF energies of 35 keV.  
This trend suggests that the spectral cutoff energy is connected to the NS magnetic field, possibly via an intermediate quantity, the most likely of which is the electron temperature, which depends on $B$ up to a saturation point.  \par

\begin{figure}[t]
\centering
\includegraphics[width=1.0\textwidth]{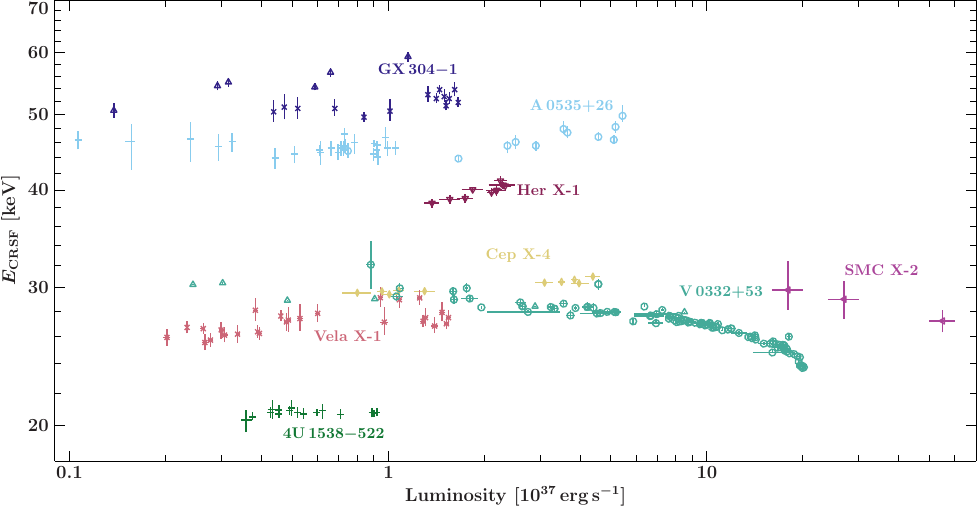}
\caption{Compilation of correlations between cyclotron line energy $E_{\mathrm{CRSF}}$ and X-ray luminosity.  Both correlations and anti-correlations can be seen.  The luminosities are calculated using the distances measured by Gaia.  Credit: \cite{staubert19}, A\&A, 622, A61, 2019, reproduced with permission \textcopyright\ ESO.}
\label{fig:staubert19}       
\end{figure}
In individual NS HMXBs, it has been observed that the CRSF line energy varies with the source X-ray flux.  As shown in Fig. \ref{fig:staubert19}, both positive correlations and negative correlations between the CRSF line energy and X-ray lumionosity have been observed \citep{staubert19}; in some cases, both trends can be observed in the same source.  
Positive CRSF energy-luminosity correlations tend to be associated with lower luminosities than negative correlations.  It has also been found that the power-law photon index ($\Gamma$) of the spectral continuum follows the opposite trend with X-ray flux as the CRSF line energy (the spectra of sources with a positive CRSF energy-flux correlation become harder with increasing flux, and vice versa).  \par
Different explanations have been suggested to explain these trends.  The positive correlation between CRSF energy and X-ray luminosity is explained either by the formation of a collisionless shock above the neutron star surface, the height of which is anti-correlated with the mass accretion rate, or by radiation pressure slowing down the relativistic plasma near the NS surface, decreasing the observed redshift of the CRSF energy as the luminosity increases.  The negative correlation observed at high luminosities is thought to be associated with a radiation-dominated shock, whose height above the NS surface increases with mass accretion rate, and the radiation from this taller accretion column illuminating a greater portion of the NS surface.  See \cite{mushtukov22} for more details about these models.

\subsubsection{4.1.2 Spectral states of Be XBs}

While most NS HMXBs do not exhibit substantial spectral variability, recently some systematic studies were performed to study the spectral states in Be XB pulsars that exhibit Type II outbursts.  Studies of the CD, HID, and power spectra of Be XB pulsars found their correlated spectral-timing behavior shares a number of similarities with LMXBs and BH-XRBs and identified two different branches in the HID during Type II outbursts (e.g. \cite{reig13}).  This behavior has been attributed to two different accretion modes which depend on whether or not the source luminosity exceeds a critical value, which is mainly determined by the magnetic field strength and is equal to $\sim (1-4) \times 10^{37}$~\ergs for the studied sources.  Given that the spectral state transitions in BH XRBs and NS LMXBs are associated with changes in their accretion disks, it is worth noting that the spectral states exhibited by these Be XBs occur during Type II outbursts, during which a transient accretion disk is expected to form around the NS.  Other examples of spectral state changes in Be XBs include GRO J1008-57 transitioning to a cold (low-ionization) disk state in between Type I outbursts, and Swift J0243.6+6124, the first Galactic ULX pulsar transitioning from a gas-supported to a radiation-supported accretion disk while experiencing a super-Eddington outburst \cite{mushtukov22}.

\subsection{4.2 Spectral states of BH systems}
\label{sec:bhstates}

\begin{figure}[t]
\centering
\includegraphics[width=0.33\textwidth, angle=90]{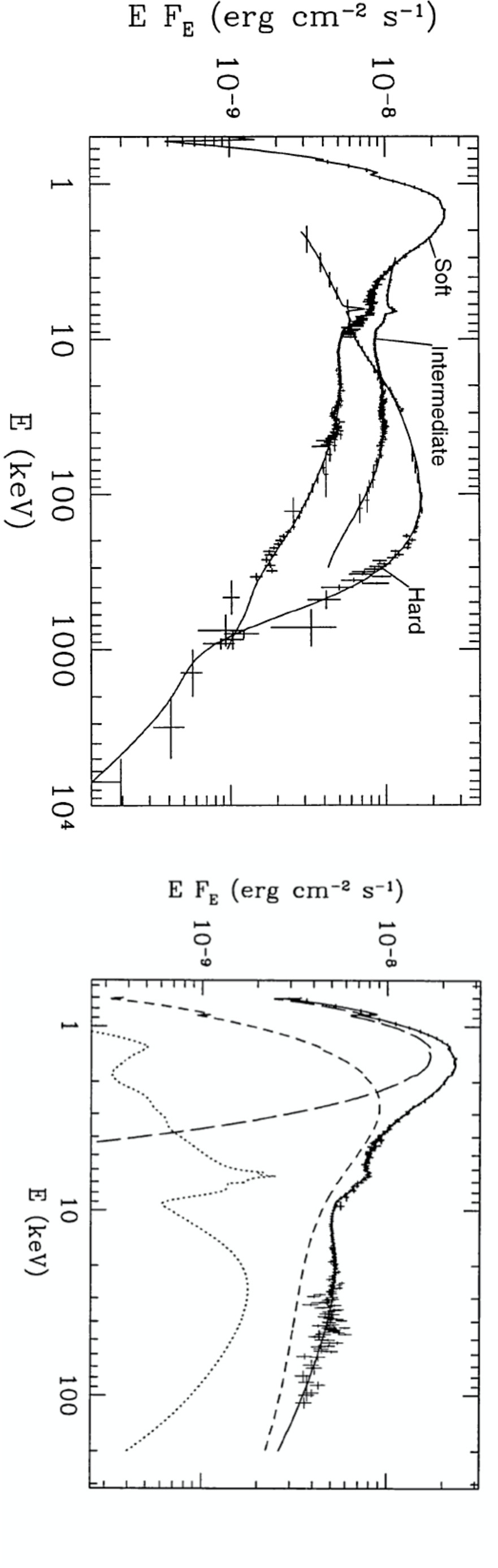}
\caption{\textit{Left:} Energy spectra of the BH HMXB Cyg X-1 in the soft, intermediate, and hard states measured by ASCA, RXTE, and CGRO/OSSE. \textit{Right:} Soft-state spectrum of Cyg X-1 from ASCA-RXTE.  Lines show the total best-fit spectral model (solid), as well as the individual model components: disk blackbody photons (long dashes), Comptonization by nonthermal electrons (dominating the short-dashed curve above the break at 15 keV) and by thermal ones (dominating the short-dashed curve below the break), and Compton reflection from the disk (dots). Credit: Figures 1 \& 7 from \cite{zdziarski00}, reproduced by permission the author.}
\label{fig:zdziarski00}       
\end{figure}

While the spectra of most NS HMXBs do not tend to show large spectral changes, BH HMXBs (e.g. Cyg X-1, Cyg X-3) transition between different \hbindex{spectral states}, as shown on the left side of Fig. \ref{fig:zdziarski00}.  
The spectrum of accreting BHs, including BH HMXBs, consists of three primary components, shown on the right side of Fig. \ref{fig:zdziarski00}: a blackbody component associated with the accretion disk, a power-law component arising from Compton up-scattering of the disk emission by non-thermal electrons in the corona, and the reflected spectrum of the coronal emission by the inner parts of the disk which exhibits an iron line and Compton hump.  Different BH spectral states are thought to result from changes to the geometry of the disk and corona associated with variations of the mass accretion rate \citep{mcclintock06}.  \par
Accreting BHs exhibit two primary spectral states referred to as the high/soft state and the low/hard state.  The high/soft state occurs at high accretion rates between about $0.1-0.5$ of the Eddington accretion rate ($\dot{M_{\mathrm{Edd}}}$). During this state, the X-ray spectrum is dominated by the thermal blackbody emission from the accretion disk, with typical temperatures of $kT_{disk}\sim1$ keV; in this state, the power-law component associated with the corona has a photon index of $\Gamma\sim2-3$ extending to MeV energies \citep{zdziarski00}.  Iron line emission arising from the inner edge of the accretion disk is especially prominent in the high/soft state.  Spectral fitting of the continuum disk emission or the relativistically broadened iron line can be used to estimate the spin of the BH \citep{mcclintock06}.  The low/hard state occurs at low accretion rates between approximately $0.01-0.1 \dot{M_{\mathrm{Edd}}}$.  In the low/hard state, the X-ray spectrum is dominated by the Comptonized continuum from the corona of electrons with a typical temperature of $kT_e\sim100$ keV and optical depth $\tau\leq 1$, which exhibits a cutoff at $50-100$ keV \citep{zdziarski00}; it is thought that in this state, the accretion disk is truncated farther away from the black hole \citep{mcclintock06}.  Radio jet emission appears in the low/hard state and disappears during the transition to the high/soft state.  For more details about the spectral states of accreting BHs, see Black holes: accretion processes.

\section{5 Variability}
HMXBs exhibit \hbindex{X-ray variability} on a wide range of timescales. 
 Some HMXBs display periodic variability.  Both orbital and super-orbital modulations in HMXB lightcurves have been observed, and NS HMXBs can exhibit X-ray pulsations associated with the rotation period of the NS.  Aperiodic variability occurs on a variety of timescales in HMXBs, from milliseconds to years.  While some HMXBs are persistent and have fairly constant luminosities that only vary a factor of a few, about 40\% of Galactic HMXBs are considered transient sources due to their large variability \citep{neumann23}, reaching faint luminosities that make them undetectable in typical surveys.  The percentage of HMXBs in the Magellanic Clouds that are transient is even higher since they are dominated by Be XBs and the majority of Be XBs are transient sources \citep{coe15,antoniou16}.

\subsection{5.1 Periodic Variability}
\subsubsection{5.1.1 X-ray pulsations}
\label{sec:pulsations}
\begin{figure}[b]
\centering
\includegraphics[width=0.9\textwidth, angle=90]{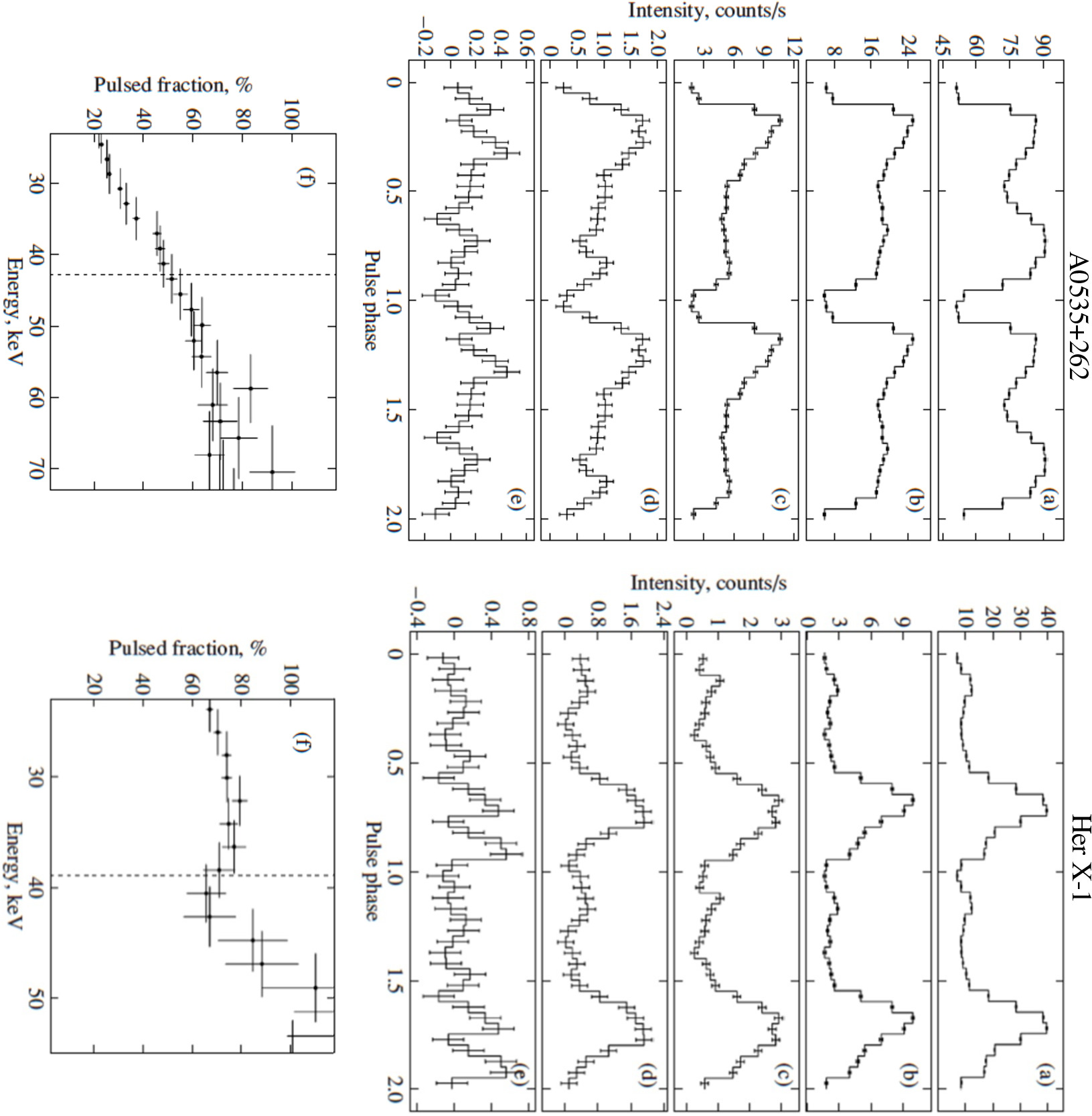}
\caption{Pulse profile for the X-ray pulsars A0535+262 and Her X-1 (in a high state) in the (a) 20–30, (b) 30–40, (c) 40–50, (d) 50–70, and (e) 70–100 keV energy bands from IBIS/INTEGRAL data. Panel (f) shows the energy dependence of the pulse profile.  The vertical dashed line indicates the cyclotron line energy of the source.  Credit: Taken from Figures 7 \& 12 from \cite{lutovinov09}.}
\label{fig:lutovinov09}       
\end{figure}
As discussed in \S\hyperref[sec:spectra]{5.1}, material accreting onto a NS in an HMXB is channeled towards the magnetic poles due to its strong magnetic field.  Therefore, as the NS rotates, \hbindex{X-ray pulsations} from emission regions near the magnetic poles can be observed and the accreting compact object is referred to as a ``pulsar". \par
The detection of X-ray pulsations and measurement of the NS spin period is performed through a periodogram or power spectrum analysis of the X-ray lightcurve.  At least half of the HMXBs in the Milky Way and the SMC have been found to be pulsars \citep{coe15, haberl16,fortin23}.  Analysis of the \chandra X-ray Visionary Program (XVP) observations of the SMC revealed that actually all \chandra sources with ${\rm L_{X}} \gtrsim 4 \times 10^{35}$\ergs\, exhibited X-ray pulsations \citep{hong17}.  The spin periods of NSs in HMXBs span a wide range from $\sim1-10^4$ seconds \citep{corbet86, reig12, townsend11}.  It has been observed that the NS spin period is anti-correlated with both the maximum soft X-ray luminosity and the average hard X-ray luminosity of the HMXB \citep{sidoli18}.  Thus, HMXBs with low NS spin periods tends to have the highest X-ray luminosities, which can be explained by the onset of the propeller effect \citep{illarionov75}; the faster the NS rotates, the greater the mass accretion rate needed to overcome the centrifugal barrier to accretion.  \par 
Once the spin period is determined, the lightcurve can be folded on the period to measure the X-ray pulse profile.  Pulse profiles often vary as a function of energy and source luminosity \citep{kretschmar19}.  For example, \cite{lutovinov09} found that some bright HMXBs exhibit double-peaked pulse profiles in the $20-40$ keV band, but one of the peaks tends to decrease in relative intensity as the energy increases making the profiles appear more single-peaked at higher energies (see left side of Fig. \ref{fig:lutovinov09}.  A feature of pulse profiles that is commonly measured is the pulse fraction, defined as $PF=(I_{\mathrm{max}}-I_{\mathrm{min}})/(I_{\mathrm{max}}+I_{\mathrm{min}})$ where $I_{\mathrm{max}}$ and $I_{\mathrm{min}}$ are the background-corrected count rates at the pulse profile maximum and minimum, respectively.  While the exact value of the pulse fraction varies from source to source, it can be as low as $10-20$\% at soft X-ray energies ($3-10$ keV) and is typically $>50$\% above 40 keV (e.g. \cite{tsygankov07,lutovinov09}.  As shown in the bottom panels of Fig. \ref{fig:lutovinov09}, in bright HMXBs, the pulse fraction tends to increase with energy, although its behavior as a function of energy is often more peculiar near the cyclotron line energy harmonics \citep{lutovinov09}.  In most cases, the pulse fraction decreases with increasing luminosity (e.g. \cite{tsygankov07, yang18}). 
\par
Variations of the pulse profile and pulse fraction are thought to be related to the geometry of the accretion flow and its changes with mass accretion rate.  At low accretion rates, the X-ray emission originates in hot spots on the NS surface and is expected to produce a ``pencil" beam of emission, as sketched in Fig. \ref{fig:hmxbschematic}.  As the accretion rate increases, a collisionless shock forms above the NS surface, and at even higher rates, a radiative shock results in the formation of an accretion column; in both of these scenarios, the chance that photons in the column will scatter off of in-falling electrons increases, causing more emission to escape out the side of the column in a ''fan" beam of emission \citep{mushtukov22}.  Different beam patterns result in different pulse profiles.  The aforementioned observed trends of the pulse fraction and pulse profiles with energy and luminosity in bright HMXBs are consistent with the presence of accretion columns in which the temperature decreases with height, resulting in hard X-rays forming in smaller regions closer to the NS surface \citep{lutovinov09}. Eclipses of parts of the accretion column, which can explain the fact that some pulse profiles of bright HMXBs transition from being double-peaked to single-peaked at higher energies \citep{lutovinov09}, can be used to estimate the NS radius and constrain the NS equation of state \citep{mushtukov18}.  par
However, pulse formation is a complex process and detailed models describing X-ray pulse profiles over a wide range of luminosity are still not fully developed.  Such models must account for gravitational light bending and for reflected emission off the NS surface from fan beam emission arising from the accretion column \citep{mushtukov22}.  The reflected emission off the NS surface also impacts the formation of CRSF, as discussed in \S\ref{sec:cyclotron}, so self-consistent models should be able to explain both the CRSF and pulse profiles.  At the super-Eddington accretion rates of ULX pulsars, the presence of an outflow can also impact the X-ray beaming and the pulse profiles \citep{mushtukov23}.
 
\subsubsection{5.1.2 Orbital periods and variability} 
\label{sec:orbital}

As discussed in \S\ref{sec:bexrb}, many Be XBs exhibit orbital variability, undergoing Type I outbursts periodically at or near periastron.  In these systems, the Type I outburst periodicity, as measured from X-ray or optical light curves, provides a way of determining the \hbindex{orbital period}.  A small number of HMXBs exhibit X-ray eclipses in their lightcurves, providing a different way of measuring the orbital period.  Orbital motion can introduce modulations in the timing of X-ray pulsations, as the pulsar moves periodically towards and away from the observer, just as orbital motion produces Doppler shifts in the optical/infrared spectra of the donor stars.  Either pulsar timing or radial velocity curves measured via spectroscopic observations provide a way of measuring both the orbital period and eccentricity of the binary system. \par
Most HMXBs follow a similar positive correlation between orbital period and eccentricity \citep{sidoli18}.  The exceptions to this trend are a few Be XBs with low eccentricity and orbital periods longer than 20 days; it is thought that these may be systems in which the NS experiences a lower natal kick \citep{townsend11}.  More eccentric systems tend to exhibit greater dynamic range in their luminosities, consistent with the fact that the compact objects in these systems experience a wider range of donor wind parameters over their orbit \citep{sidoli18}, although eccentricity is not the only factor that can result in a high dynamic range.  \par
\begin{figure}[b]
\centering
\includegraphics[width=0.8\textwidth]{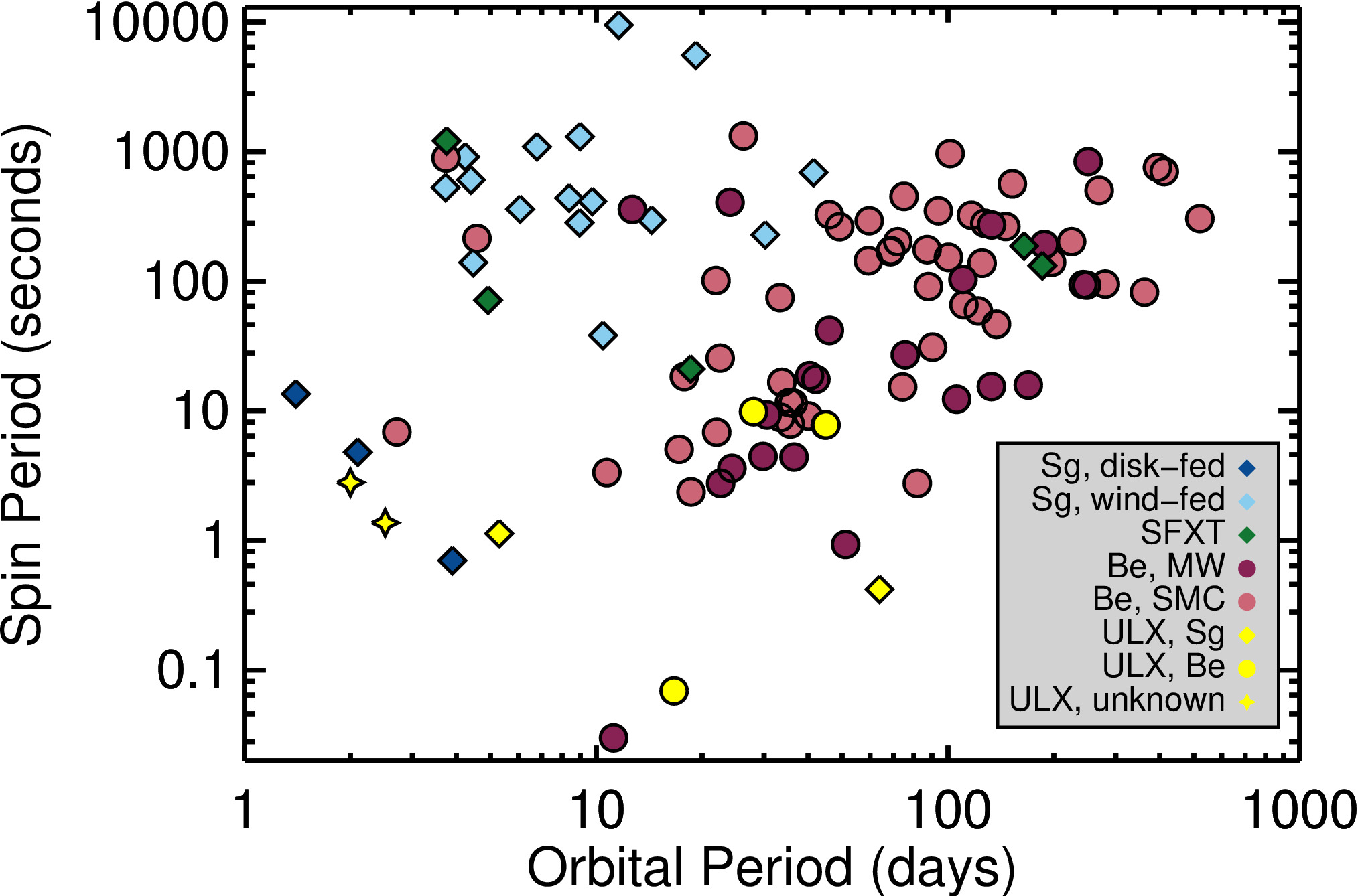}
\caption{The spin period versus orbital period of different groups of HMXBs are shown with different symbols, as indicated in the legend.  Data is compiled from \cite{corbet86, grebenev09, townsend11, reig12, klus14, coe15, martinez17, king19, townsend20}.}
\label{fig:corbet}       
\end{figure}

Different classes of HMXBs occupy different parts of the spin-orbital period parameter space (e.g. \cite{corbet86, townsend11, sidoli18}), which is often referred to as the Corbet diagram.  A compilation of the spin and orbital periods of different types of HMXBs is shown in Fig. \ref{fig:corbet}.  Disk-fed HMXBs, including some ULXs \citep{townsend20}, have the shortest orbital periods ($\lesssim3$ days) and fast spins; their spin and orbital periods may be anti-correlated, but there are too few such sources with spin and orbit measurements to make a definitive conclusion.  The fast rotations of the NSs in these systems are thought to result from efficient spin-up by the accretion disk. \par
Classical wind-fed Sg XBs have high spin periods and orbital periods of $3-60$ days, in between those of disk-fed HMXBs and Be XBs.  No clear correlation is seen between the spin and orbital periods of wind-fed Sg XBs.  This has often been interpreted as a result of inefficient angular momentum transfer by wind accretion.  The orbital periods of SFXTs span a wide range, overlapping with both Sg and Be HMXBs, although they tend to have higher eccentricities at a given orbital period than other HMXBs \citep{sidoli18}.  \par
The vast majority of Be XBs have longer orbital periods than Sg XBs and lower average spin periods.  The spin and orbital periods of Be XBs are positively correlated, with slower spinning pulsars residing in longer period orbits, suggesting that the NSs in Be XBs evolve towards an equilibrium period \citep{corbet86}.  Angular momentum can be efficiently transferred from the Be equatorial decretion disk to the compact object during periastron passage \citep{waters89}; if the angular velocity of the infalling material is greater than that of the magnetosphere, it is accreted and spins ups the NS, and otherwise the propeller effect flings the infalling material away, spinning down the NS in the process \citep{illarionov75}. \par
Using the latest census of Be XBs in the Milky Way and the Magellanic Clouds, two subpopulations with different characteristic spin and orbital periods and orbital eccentricities have been identified \citep{knigge11}.  The bimodality is more prominent in the log\pspin\, distribution (than in the log\porb), with the two peaks shown at \pspin$\sim$10 s (\porb$\sim$40 d) and \pspin$\sim$200 s (\porb$\sim$100 d), respectively.  Be XB pulsars with \pspin$<$40 s were found to more likely experience type II outbursts and exhibit greater long-term X-ray variability compared with those with \pspin$>$40 s \citep{haberl16}; these trends suggest that the \pspin\, bimodal distribution is likely due to different accretion modes in Be XBs \citep{haberl16}, although it has also been attributed to the two types of NS-forming supernovae, electron-capture versus iron-core-collapse supernovae \citep{knigge11}. 

\subsubsection{5.1.3 Superorbital Modulations}
\label{sec:superorbital}
Superorbital modulations have been observed in the X-ray lightcurves of several HMXBs.  These modulations are periodic or quasi-periodic luminosity variations on timescales longer than the orbital period, typically by a factor of $3-10$ (e.g. \cite{kotze12} and references therein).  The origins of superorbital modulations vary for different types of HMXBs.  In disk-fed HMXBs, such as SMC X-1, these modulations are thought to result from the precession of a tilted or warped accretion disk periodically obscuring the X-ray source or varying the geometry of hot spots in the accretion flow \citep{kretschmar19}.  Be XBs exhibit both X-ray and optical superorbital variability.  The optical modulations are likely associated with the formation and depletion of the Be star's circumstellar disks or with the neutron star's orbit impacting the precession or warping of this circumstellar disk \citep{rajoelimanana11}.  These variations in the circumstellar disk can modulate the accretion onto the compact object, resulting in X-ray superorbital periods.  ULXs and disk-fed HMXBs both follow the same correlation between superorbital and orbital periods, so it has been suggested that the superorbital periods in ULXs may similarly be attributed to the modulation of precessing hot spots or density waves in the accretion or circumstellar disk by the binary motion of the system \citep{townsend20}. \par
Several theories for the origin of superorbital modulations in wind-fed HMXBs have been proposed (see \cite{kretschmar19} and references therein).  This variability may be driven by tidally-regulated oscillations of the outer layers of the supergiant donor resulting in a variable accretion rate onto the compact object.  Another possibility is that these variations in the accretion rate may result from interaction between the neutron star and co-rotating interaction regions (CIRs) of the supergiant, which are spiral-shaped density and velocity perturbations in the stellar wind.  In some cases, for example in 4U 1820-30, the superorbital period may be caused by the presence of a third stellar companion inducing an eccentricity in the inner binary, which in turn modulates the mass transfer rate.  More observations spanning many superorbital cycles are required for more wind-fed SgHMXBs to better understand the origin of their superorbital modulations.

\subsection{5.2 Aperiodic Variability}
\subsection{5.2.1 Short-Timescale Variability}

On timescales of milliseconds to hours, HMXBs exhibit aperiodic variability which produces a red noise power-law continuum in power spectra of their X-ray lightcurves \citep{belloni90}. The power spectrum of some HMXBs demonstrates a break in their power-law continuum.
For most accreting X-ray pulsars as well as accreting black holes, the red noise power-law index is $\alpha\approx1.4-2.0$ at frequencies above the break and $\alpha\approx0-1$ at frequencies below the break \citep{hoshino93}; for accreting pulsars, this break frequency is close or equal to the pulsation frequency, and for Roche-lobe overflow systems, it is close to the orbital frequency \citep{icdem11}.  Several other hydrodynamics processes have also been suggested to contribute to the aperiodic variability, including: (1) density fluctuations due to magnetohydrodynamic turbulence in the accreting plasma \citep{hoshino93}, (2) stochastic perturbations in the accretion disk being advected to the magnetospheric radius \citep{revnivtsev09}, (3) Rayleigh-Taylor instabilities occurring at the magnetospheric boundary when the accretion rate is low and the flow is subsonic \citep{shakura13}, and (4) wind inhomogeneities or instabilities in the shock front resulting from the photo-ionization of the stellar wind by the NS \citep{manousakis15}.  \par
The power spectra of HMXBs can exhibit quasi-periodic oscillations (QPOs), which tend to be well-fit by Lorentzian functions and whose physical origin, although still not completely understood, is thought to be associated with processes or inhomogeneities in the inner accretion disk (e.g. \cite{kaur08,ingram19}).  QPOs have been observed in both BH and NS HMXBs; QPOs in NS HMXBs can provide constraints on the magnetic field strength of the NS since the accretion disks in NS HMXBs, if they are present, are truncated at the magnetosphere.  Finally, in the case of NS HMXBs, pulsations of the X-ray emission originating in the vicinity of the magnetic poles of the NS can introduce peaks in the power spectrum at the NS spin frequency and its harmonics \citep{belloni90}.  Thus, the power spectra of HMXBs can used to probe important properties of the compact objects and the accretion flow.

\subsection{5.2.2 Long-Timescale Variability}
\begin{figure}[b]
\centering
\includegraphics[width=0.8\textwidth]{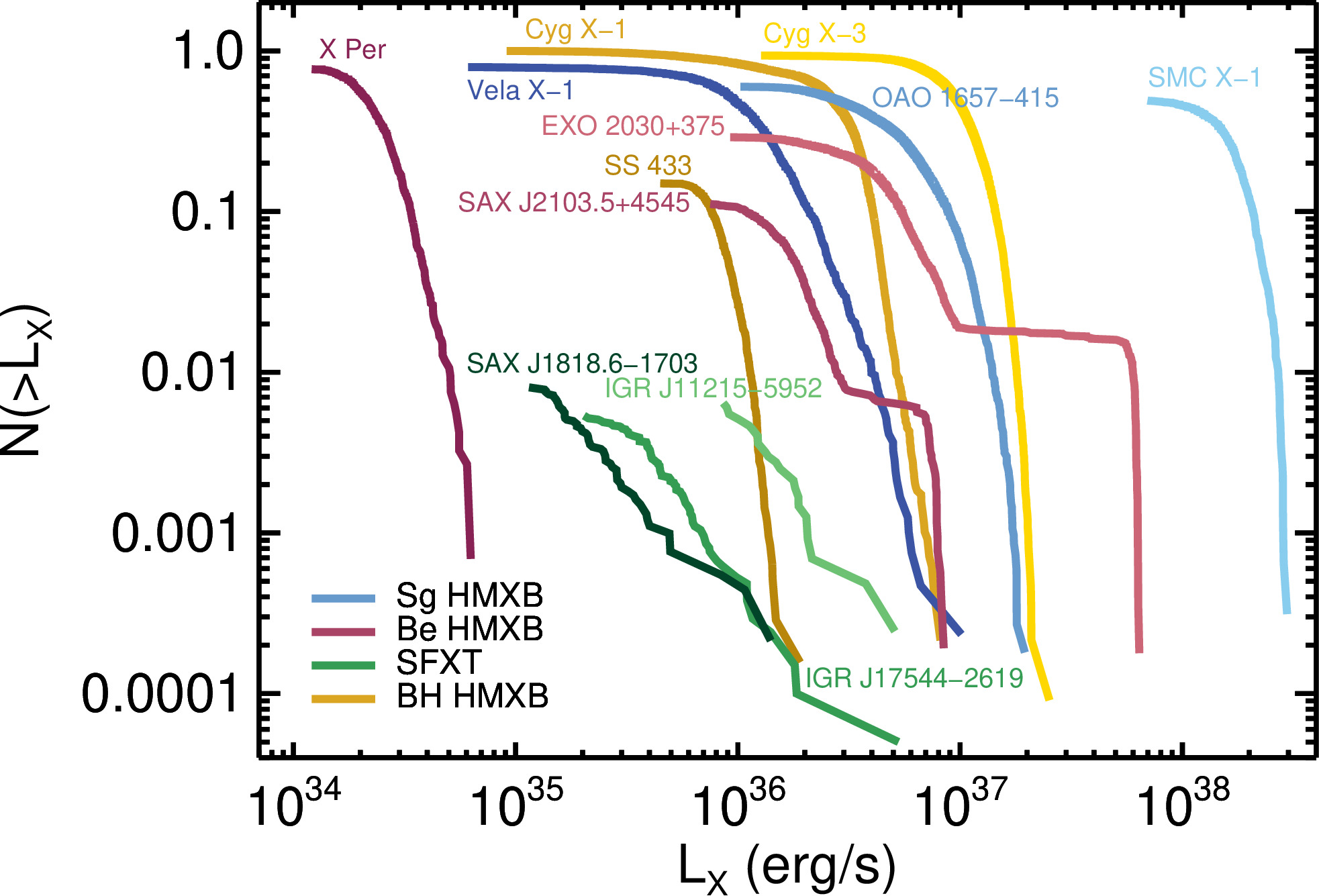}
\caption{Cumulative 18-50 keV luminosity functions of a representative set of HMXBs measured from \textit{INTEGRAL} observations divided into 2 kilosecond bins.  Different classes of HMXBs are shown in different color shades as indicated in the legend. The HMXB names are written next to the curves in the corresponding color.  Credit: Adapted from Figures 1-4 of \cite{sidoli18}, ``An INTEGRAL overview of High-Mass X-ray Binaries: classes or transitions?", MNRAS, 481, 2.}
\label{fig:sidoli18}       
\end{figure}
On longer timescales of $~$days to years, different classes of HMXBs can exhibit different types of variability, including the periodic variability described in the previous section and aperiodic variability including the ``off-states" and the flaring behavior of some Sg XBs discussed in \S\hyperref[sec:sghmxb]{3.1}.  A useful way to encapsulate and visualize the variety of long-term behavior displayed by HMXBs is with cumulative luminosity distributions (CLDs), which measure the fraction of time that a source is observed above a given luminosity.  The biggest challenge in measuring CLDs that provide an accurate representation of the source variability is that they require monitoring over long timescales.  Even when based on long monitoring campaigns, it is important to be mindful of the fact that CLDs can be uncertain at the high luminosity end due to potentially missing rare, high-luminosity events, and uncertain at the low luminosity end due to the incompleteness of detections when the source flux nears the sensitivity of the instrument.  From a CLD diagram, it is straightforward to determine key variability metrics.  The maximum value along the y-axis that a source reaches in a CLD diagram is referred to as the duty cycle, the fraction of the observed time that the source is detected above the minimum luminosity set by the sensitivity of the instrument.  The variability amplitude of the source is shown by the spread of values along the x-axis. \par
The most comprehensive studies of HMXB CLDs to date are reported in \cite{bozzo15} and \cite{sidoli18}, which focus on the soft and hard X-ray bands, respectively.  \cite{bozzo15} present the $2-10$ keV CLDs of 11 HMXBs based on monitoring campaigns lasting from approximately 2 weeks to 2 years with \textit{Swift} XRT; three of the sources studied were classical Sg XBs, six were SFXTs, and two were ``intermediate'' SFXTs, with X-ray properties in between those of classical HMXBs and SFXTs.  The $18-50$ keV CLDs studied in \cite{sidoli18} are based on 14 years of monitoring observations by \textit{INTEGRAL} IBIS/ISGRI; this work includes 58 HMXBs, the vast majority of which reside in the Milky Way and include both Sg and Be HMXBs.  A representative sample of the $18-50$ keV CLDs from \cite{sidoli18} are shown in Fig. \ref{fig:sidoli18}.  \par
At both soft and hard X-ray energies, the CLDs of classical Sg XBs exhibit a single knee \citep{bozzo15,sidoli18}.  Compared to classical systems, the soft and hard X-ray CLDs of SFXTs are shifted to lower luminosities by factors of $10-100$ at comparable duty cycles and they exhibit large variability amplitudes of $>100$ compared to $<40$ for classical systems.  In the $2-10$ keV band, the CLDs of some SFXTs display plateaus, but these features are not seen in the $18-50$ keV CLDs; it is not clear whether these plateaus are an artifact resulting from the more limited monitoring campaigns by \textit{Swift} compared to \textit{INTEGRAL} or if they are real features possibly related to the triggering mechanisms responsible for the highest accretion rates in SFXTs \citep{bozzo15}.  Given the larger HMXB sample in the study by \cite{sidoli18}, some additional trends can be noted for the hard X-ray CLDs.  The CLDs of classical Sg XBs follow a log-normal distribution while those of SFXTs are power-law distributions with a power-law slope of $\approx2$.  Among the classical Sg XBs, the RLO systems reach the highest luminosities, and the CLDs of sources with higher median luminosities and lower spin periods tend to be steeper with a lower variability amplitude than those of sources with lower median luminosities.  The $18-50$ keV CLDs of Be XBs have high variability amplitude ($>100$), higher average hard X-ray luminosity in outburst than SFXTs, and duty cycles that are intermediate between those of classical Sg XBs and SFXTs.  They also exhibit much more complex shapes than any of the Sg XBs, often with multiple knees, indicative of the different types of outbursts (Type I and Type II) that they can experience or different accretion regimes.  CLDs thus provide a way of distinguishing between different classes of HMXBs and of identifying HMXBs with intermediate properties based on their long-term variability.  
\section{6 HMXB Populations in the Milky Way and Magellanic Clouds}

Due to the \hbindex{Milky Way}'s large footprint on the sky and the transient nature of many HMXBs, our understanding of the Galactic HMXB population has largely depended upon the identification of HMXBs among the X-ray sources detected and monitored over time by all-sky surveys and other large-scale surveys.   
Due to the high levels of obscuration that exist near the Galactic plane, hard X-ray surveys have been critical for gathering as complete a census as possible of the Milky Way's HMXB population.  \textit{INTEGRAL}'s dedicated monitoring survey of the Galactic Plane has made particularly significant contributions to our understanding of Galactic HMXBs, reaching sensitivity of $7\times10^{-12}$ erg s$^{-1}$ cm$^{-2}$ in the 17-60 keV band over half of the Plane at latitude $|b|<17.5^{\circ}$ \citep{krivonos12} and discovering highly obscured HMXBs as well as SFXTs.  
The most recent catalogs of HMXBs in the Milky Way can be found in \cite{liu06, walter15, fortin23, neumann23} and contain approximately 150 unique confirmed and candidate HMXBs.  About a third of Galactic HMXBs are classified as Sg XBs (including SFXTs), $\approx$50\% are Be HMXBs, and the donor star remains unclassified in about 20\% \citep{fortin23}.  

The \hbindex{Magellanic Clouds} have been excellent targets for studies of young XRB populations at a depth similar to that of Galactic studies, but without the limitations of distance uncertainties and high obscuration.  The Small Magellanic Cloud (SMC), our second nearest 
star-forming galaxy at a distance of 62 kpc, harbors an HMXB population comparable to the observed population in the Milky Way (e.g., \cite{liu05,antoniou10,coe15,haberl16}). 
By combining data from \cite{haberl16} and \cite{antoniou19}, 137 confirmed and highly likely HMXBs have been identified. The spectroscopic properties of these sources (or, when not available, their multi-wavelength photometric properties) have allowed the classification of the vast majority as Be XBs (e.g., \cite{antoniou09,coe15}). 

In contrast, due mainly to its large extent on the sky and large-scale diffuse X-ray emission from the hot ISM, our nearest neighbor at 50 kpc, the Large Magellanic Cloud (LMC) has been surveyed in the X-rays less extensively and to lower sensitivity limits than the SMC. 
Thus, the number of known HMXBs in the LMC remains significantly smaller than that of the SMC.  \cite{antoniou16} compiled a list of 40 confirmed and candidate HMXBs, and classified them based on their optical photometric properties.  Subsequently, two additional studies \citep{vanjaarsveld18,haberl22} increased the known HMXB population of the LMC to $\sim$60 members (25 XRB pulsars and $\sim$35 candidate HMXBs from the literature).  The bulk of the LMC HMXBs have been classified as Be or candidate Be HMXBs, with only four LMC HMBs being Sg or candidate Sg XBs \citep{antoniou16, vanjaarsveld18}.  In the recent future, deeper X-ray observations from the \textit{Chandra} Very Large Program survey of the LMC will reach comparable sensitivities to X-ray surveys of the SMC, enabling the discovery of fainter HMXBs in this galaxy.  Furthermore, \textit{eRosita}'s all-sky surveys, given their high cadence and and sensitivity, are likely to discover new HMXBs in both the Galaxy and Magellanic Clouds. 

\subsection{6.1 HMXB Luminosity function}

X-ray binaries are the primary sources of X-ray emission in ``normal" galaxies without an active galactic nucleus (AGN), with HMXBs being the dominant source in star-forming galaxies (e.g. \cite{mineo12}).  In addition to being a key ingredient to the X-ray emission of galaxies, the \hbindex{X-ray luminosity function} (XLF) of HMXBs also can provide some insights into the accretion processes occurring in these systems.  \par
The XLF of Galactic HMXBs is challenging to measure.  It requires X-ray surveys of fairly uniform sensitivity along the Galactic plane, and accurate distance measurements to individual HMXBs.  Furthermore, it is necessary to correct for the fraction of the HMXB population residing in regions of the Milky Way (MW) that are not observable given the sensitivity limit of the X-ray survey, which requires modeling the mass distribution of the Galaxy and its relationship to the spatial distribution of HMXBs.  \par
\begin{figure}[t]
\centering
\includegraphics[width=0.7\textwidth]{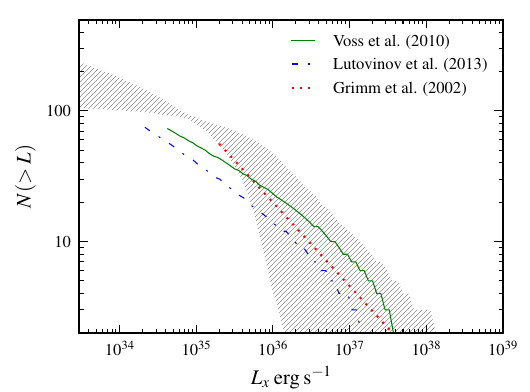}
\caption{The shaded area represents the cumulative HMXB luminosity function (1-200 keV) as reconstructed by \cite{doroshenko14} through modeling the observed flux distribution of HMXBs detected by the \textit{INTEGRAL} Galactic survey.  Best fits of the HMXB luminosity function from previous studies converted to the $1-200$ keV are also shown.  Credit: Taken from Figure 2 from \cite{doroshenko14}, A\&A, 567, A7, 2014, reproduced with permission \textcopyright\ ESO}.
\label{fig:doroshenko14}       
\end{figure}

Despite these challenges, the Galactic HMXB XLF measurements made by different studies are in general agreement with one another, as shown in Fig. \ref{fig:doroshenko14}.  Between X-ray luminosities of $L_X\sim10^{35}-10^{37}$ erg s$^{-1}$, the XLF is well-fit by a power-law ($dN/dL \propto L^{-\alpha}$) with best-fitting index $\alpha\approx1.3-1.7$ (see \cite{lutovinov13} and references therein).  The Galactic XLF exhibits a break or cutoff around $L_X\sim10^{37}$ erg s$^{-1}$, and its power-law slope at higher luminosities is $\alpha>2$.
Such high luminosities would require near Eddington or super Eddington accretion rates, which likely can only be reached as HMXBs approach RLO; thus, the relative scarcity of high luminosity sources is consistent with the fact that RLO is unstable in HMXBs and therefore short-lived (see \S\ref{sec:accretion}).  The Galactic XLF measurements extending to the faintest luminosities ($L_X\sim10^{34}$ erg s$^{-1}$) show a hint of flattening below $10^{34}-10^{35}$ erg s$^{-1}$ (\cite{lutovinov13, doroshenko14}).  A faint-end flattening of the HMXB XLF also seems to be supported by recent results \citep{tomsick17, clavel19} based on \textit{NuSTAR} observations that find a dearth of low-luminosity HMXBs compared to expectations based on the XLF measured by \cite{lutovinov13}.  This flattening can, at least partly, be attributed to the propeller effect, the centrifugal inhibition of accretion due to the interaction of the accretion flow with the pulsar’s magnetic field \citep{illarionov75}. 
Both the number of known MW HMXBs and the number of HMXB distance measurements has increased by $\approx$50\% since these studies of the Galactic HMXB XLF were carried out, thanks in large part to INTEGRAL's ongoing Galactic Plane survey and parallax measurements from \textit{Gaia}, so tighter constraints on the Galactic HMXB XLF are possible in the near future. \par
The shape of the XLF of MW HMXBs is consistent with the universal XLF of HMXBs in the range $L_{\mathrm{2-10 keV}}\approx10^{35}-10^{37}$ erg s$^{-1}$.  In the 2-10 keV band, the XLF of HMXBs in other star-forming galaxies exhibits a consistent power-law slope of $\alpha\approx1.6$ between $10^{35}-10^{40}$ erg s$^{-1}$ and a cutoff at $L_X\sim10^{40}$ erg s$^{-1}$ (e.g. \cite{mineo12}).  Thus, the MW XLF appears to have a lower-luminosity cutoff compared to other star-forming galaxies, but its power-law slope below that cutoff is consistent with the universal XLF.  Unlike the slope, the normalization of the HMXB XLF varies from galaxy to galaxy because it depends on the galaxy's star formation rate (SFR) (e.g. \cite{mineo12}).  \par
The XLFs of HMXBs in the LMC and SMC are also consistent with the power-law slope of the universal XLF between $L_X\approx10^{35}-10^{37}$ erg s$^{-1}$, although these XLFs were based on small numbers of sources \citep{shtykovskiy05a,shtykovskiy05b}.  Both of these XLFs appear to flatten below $\sim10^{35}$~\ergs, similar to the observed hint of flattening in the MW XLF that may be due to the propeller effect \citep{tomsick17, clavel19}.  Tighter constraints on the LMC and SMC XLFs can now be made using the larger HMXB samples provided in the most recent catalogs.

\subsection{6.2 Spatial distribution and ages}
Our position within the Milky Way's spiral disk gives us an edge-on view of our Galaxy that allows us to measure the vertical scale height of HMXBs, but makes it more challenging to measure their radial distribution as it is more difficult to detect and determine accurate distances to HMXBs on the far side of the Galaxy due to obscuration by dust and smaller parallexes.  The HMXB scale height has been found to be $\approx100$ pc, which is larger than the scale heights of massive stars \citep{lutovinov13}.
This result suggests that HMXBs have traveled farther from the Galactic plane than their parent stellar populations due to supernova kicks imparted to the binary systems when the compact object is formed.  Using a value of $100$ km s$^{-1}$ for the typical systemic velocity of HMXBs (e.g. \cite{bodaghee12}), the typical vertical distance traveled by HMXBs from their parent populations can be used to place a lower limit\footnote{This estimate is a lower limit given that only the vertical distance traveled, which is a 1D projection of the binary system's 3D motion, has been considered.} on their kinematic ages of $\approx1$ Myr.  Kinematic ages provide an estimate of the typical time lag between the formation of the first compact object when the more massive star explodes as a supernova and the onset of the X-ray bright, accretion phase. \par
\begin{figure}[t]
\centering
\includegraphics[width=0.7\textwidth]{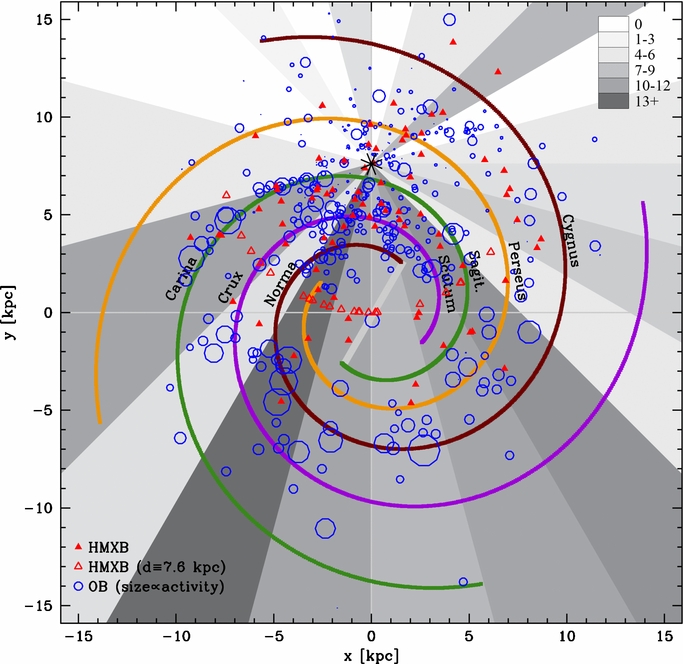}
\caption{Galactic distribution of HMXBs with known distances as of 2012 (79, filled triangles) and the locations of OB associations (458, circles)). The symbol size of the latter is proportional to the amount of activity in the association (amount of ionizing photons as determined from the radio continuum flux). Galactic spiral arm model is overlaid with the Sun situated at 7.6 kpc from the Galactic center (GC). HMXBs whose distances are not known have been placed at 7.6 kpc (23, empty triangles), i.e., the Sun–GC distance. The shaded histogram represents the number of HMXBs in each 15 deg bin of galactic longitude as viewed from the Sun.  Credit: Figure 2 from \cite{bodaghee12}, reproduced with permission \textcopyright AAS.}
\label{fig:bodaghee12}       
\end{figure}
The radial distribution of HMXBs in the Milky Way is found to be similar to the distribution of giant HII regions and CO gas associated with the spiral arms (e.g. \cite{lutovinov13}), a result consistent with the expected young ages (few to tens of Myr) of HMXBs (e.g. \cite{linden10}).  Correcting the observed distribution of HMXBs for INTEGRAL's survey sensitivity, \cite{lutovinov13} found that it peaks between $2-8$ kpc from the Galactic Center and is consistent with the star formation rate surface density.  Other works (including \cite{bodaghee12} shown in Fig. \ref{fig:bodaghee12}), using larger samples of HMXBs calculated the spatial cross-correlation function between HMXBs and OB associations or star forming complexes, finding significant clustering and average offsets of $0.4\pm0.2$ kpc \citep{bodaghee12} and $0.3\pm0.05$ kpc \citep{coleiro13}.  The observed pattern of offsets in these studies cannot be explained by Galactic rotation alone, but are consistent with natal kick velocities of $100\pm50$ km s$^{-1}$ and kinematic ages of $\approx4$ Myr. \par 
The total ages of HMXBs, not just their kinematic ages, can be determined by associating HMXBs with their birthplaces, which requires accurate distances.  \cite{coleiro13} estimated the ages of a small sample of 13 HMXBs by first obtaining more precise distance estimates through the fitting of their optical to near-infrared spectral energy distribution; they then compared the HMXB positions to the locations of the Galactic spiral arms, and modeled the rotation of the arms and the motion of HMXBs due to Galactic rotation alone to determine the point in time when the position of each HMXB most closely coincided with one of the spiral arms.  With this method, they found mean ages of $\sim45$ Myr for Sg XBs and $\sim51$ Myr for Be HMXBs.  While these estimates are consistent with the different evolutionary timescales expected for these two classes of HMXBs, these results are not conclusive due to the small sample size.  With new parallax distance estimates to many HMXBs from \textit{Gaia} \citep{fortin23}, it should be possible to improve measurements of Galactic HMXB ages. \par 
Studies of the spatial distribution and ages of HMXB can be more easily carried out in the Magellanic Clouds.  By comparing the positions of HMXBs in the SMC and LMC with spatially resolved star formation history (SFH) maps, it has been determined that the Be XBs in the SMC are associated with bursts of star formation that occurred $\sim20-60$ Myr ago (e.g. \cite{antoniou10}), while LMC HMXBs are associated with a more recent star formation episode ($\sim$6-40 Myr ago) (e.g. \cite{antoniou16}).  The spatial cross-correlation function between HMXBs and OB associations in the SMC reveals strong clustering \citep{bodaghee21}.  The average distance between an HMXB and its nearest OB association in the SMC is 150 pc; this offset is lower in the SMC Bar ($120\pm90$ pc) than in the SMC Wing ($450\pm180$ pc), which faces the LMC.  Since a large fraction of the Be XBs in the SMC are connected with a burst of star formation $\approx40$ Myr ago, and stellar evolution models predict it takes $\approx5-20$ Myr for a high-mass star to undergo a supernova explosion and collapse into a compact object, the period of time that the HMXBs have had to migrate away from their OB associations of origin as a result of supernova kicks is $\approx20-35$ Myr.  From this estimate, it can be calculated that the kick velocities received by SMC HMXBs are typically $2-34$ km s$^{-1}$ \citep{bodaghee21}, much lower than those of their Milky Way counterparts, in agreement with previous studies.
While the transverse velocities of Galactic Be HMXBs are lower by a factor of $\approx3$ compared to Galactic Sg XBs (which can be explained by differences in their evolution \citep{vandenheuvel00}), the difference between the Galactic and SMC HMXB space velocities cannot be fully ascribed to the overabundance of Be HMXBs in the SMC \citep{bodaghee21}.  Similar studies of the spatial distribution of HMXBs in the LMC have yet to be carried out. 

\subsection{6.3 Comparing the Milky Way and Magellanic HMXB populations}
\label{sec:comparison}

As mentioned in previous sections, there are significant differences between the observed HMXB populations in the Milky Way and Magellanic Clouds.  While roughly a third of MW HMXBs are now known to host supergiant donors and 50\% host Be stars \citep{fortin23}, in the SMC, only two HMXBs, SMC X-1 and CXOU J005409.57-724143.5, have been classified as Sg XBs compared to the $\approx70$ that have been classified as Be HXMBs (\cite{coe15, haberl16}).  While the LMC HMXB population has not been as thoroughly studied as that of the SMC, of the $\approx30$ HMXBs that have been classified, only 4 are confirmed or candidate Sg XBs (\cite{antoniou16, vanjaarsveld18}).  Thus, the HMXB populations of both Magellanic Clouds, especially the SMC, appear to have a greater ratio of Be/Sg XBs than the Milky Way.  It is unlikely that a significant population of Sg XBs in the Magellanic Clouds remain undetected due to obscuration given their low dust content; furthermore, INTEGRAL surveys of the SMC and LMC have not uncovered highly obscured Sg XBs as they did in the Milky Way (\cite{coe10,grebenev13}) and monitoring campaigns of the SMC have discovered transient Be XBs but no SFXTs \citep{kennea18}.  \par
The overabundance of Be XBs in the SMC has been noted and investigated by several studies.  The Be XB population of the SMC was estimated to be larger than that of the Milky Way by a factor of $\approx1.5$ after accounting for differences in their star formation rates (\cite{antoniou09, shtykovskiy05a}).  The HMXB formation efficiency of the SMC was found to increase as a function of time following a burst of star formation up to $\sim$40--60 Myr, reaching a peak efficiency that is higher than that of the LMC by a factor of $\approx17$ (\cite{antoniou16, antoniou19}).  One factor that is thought to contribute to the abundance of HMXBs in the SMC is its low metallicity (Z$\sim$0.2\zsun); lower metallicity stars have weaker stellar winds, which impacts their evolution and is predicted to result in more numerous and more luminous HMXB populations (e.g. \cite{linden10}).  \par
The first theoretical study that addressed the effect of metallicity on the formation and evolution of HMXBs and compared its results to the HMXB populations of the Milky Way and Magellanic Clouds was that of \cite{dray06}.  This study performed extensive Monte Carlo simulations of binary systems that were able to reproduce the orbital properties and X-ray luminosities of Galactic HMXBs and found that the number of HMXBs increased with decreasing metallicity.  The simulation results for half solar metallicity were in decent agreement with the properties of LMC HMXBs, although the LMC comparison sample was small.  However, while the predicted increase in the number of HMXBs by a factor of 3 for the subsolar metallicity environment of the SMC could be consistent with the observed number of SMC HMXBs, the orbital period of the distribution of the simulated binaries was skewed to much lower orbital periods than the observed distribution.  This study showed that metallicity alone cannot explain the observed SMC HMXB population; in this work, the orbital period distribution of the SMC HMXBs could only be reproduced by an HMXB population associated with a very large burst of star formation $\sim30-100$ Myr ago, consistent with the estimated ages of the SMC HMXBs by observational studies (e.g. \cite{antoniou10}).  \par
Recently, it has been shown that the shape of the high-luminosity end of the HMXB XLF depends on the metallicity of the young stellar population; while the HMXB XLF has the same power-law slope independent of metallicity at $L_X<(3-10)\times10^{37}$ erg s$^{-1}$, but above $10^{38}$ erg s$^{-1}$, the XLF gradually flattens and extends to higher luminosities with decreasing metallicities \citep{lehmer21}.  This trend results from formation of a greater number of luminous HMXBs, including ULXs, at lower metallicities (e.g. \cite{dray06, linden10, fragos13}).
The bright end of the SMC XLF above $10^{37}$ erg s$^{-1}$ appears to flatten in a way that is consistent with the XLFs of other low-metallicity galaxies \citep{lehmer21}.  \par
Studies of HMXBs in the Milky Way and Magellanic Clouds have thus demonstrated that both metallicity and age impact the properties of HMXB populations.  These effects are being investigated in more detail by both theoretical simulations (e.g.\cite{linden10, fragos13}) and observational studies of HMXBs in other galaxies (e.g. \cite{lehmer21}).  Studying the impact of both metallicity and age on HMXB properties is relevant to understanding the gravitational wave sources that may evolve from HMXBs as well as the contribution of the first generations of HMXBs in the  Universe to the heating of intergalactic gas during the Epoch of Reionization.

\bibliographystyle{aps-nameyear}  
\bibliography{refs}

\end{document}